\documentclass[10pt,a4paper]{article}

\usepackage{caption}
\usepackage{subcaption}
\usepackage{graphicx}
\graphicspath{{figures/}}
\usepackage[dvipsnames]{xcolor}
\usepackage{xcolor}
\usepackage{framed}

\usepackage{amssymb}
\usepackage{amsmath} 

\usepackage{algorithmic}
\usepackage{array}
\usepackage{lipsum}
\usepackage{hyperref}
\usepackage{cite}

\begin{document}

\title{\textbf{A stylised view on structural and functional connectivity in dynamical processes in networks}}

\author{
Venetia Voutsa$^{1}$, 
Demian Battaglia$^{2, 3}$,
Louise J. Bracken$^{4}$, 
Andrea Brovelli$^{5}$, \\ 
Julia Costescu$^{4}$, 
Mario Díaz Muñoz$^{6}$, 
Brian D. Fath$^{7, 8, 9}$, 
Andrea Funk$^{10, 11}$,\\ 
Mel Guirro$^{4}$,
Thomas Hein$^{10, 11}$, 
Christian Kerschner$^{6, 9}$,
Christian Kimmich$^{9, 12}$, \\
Vinicius Lima$^{2, 5}$, 
Arnaud Mess\'e$^{13}$,
Anthony J. Parsons$^{14}$, 
John Perez$^{5}$,\\
Ronald P\"oppl$^{15}$, 
Christina Prell$^{16}$,
Sonia Recinos$^{10}$,
Yanhua Shi$^{9}$, 
Shubham Tiwari$^{4}$, \\
Laura Turnbull$^{4}$, 
John Wainwright$^{4}$, 
Harald Waxenecker$^{9}$, 
and Marc-Thorsten H{\"u}tt$^{1}$ }

\date{
\small
$^{1}$ Department of Life Sciences and Chemistry, Jacobs University Bremen, 28759 Bremen, Germany \\
$^{2}$ Aix-Marseille Université, Inserm, Institut de Neurosciences des Systèmes (UMR 1106), Marseille, France\\
$^{3}$ University of Strasbourg Institute for Advanced Studies (USIAS), Strasbourg, France\\
$^{4}$ Department of Geography, Durham University, DH1 3LE, Durham, UK \\
$^{5}$ Institut de Neurosciences de la Timone UMR 7289, Aix Marseille Universit\'e, CNRS,
13005, Marseille, France \\
$^{6}$ Department of Sustainability, Governance and Methods, Modul University Vienna, 1190 Vienna, Austria\\
$^{7}$ Department of Biological Sciences, Towson University, Towson, Maryland 21252, United States \\
$^{8}$ Advancing Systems Analysis Program, International Institute for Applied Systems Analysis, Laxenburg A-2361, Austria\\
$^{9}$ Department of Environmental Studies, Masaryk University, 60200 Brno, Czech Republic\\
$^{10}$ Institute of Hydrobiology and Aquatic Ecosystem Management (IHG), University of Natural Resources and Life Sciences Vienna (BOKU), 1180, Vienna, Austria\\
$^{11}$ WasserCluster Lunz - Biologische Station GmbH, Dr. Carl Kupelwieser Promenade 5, 3293 Lunz am See, Austria\\
$^{12}$ Regional Science and Environmental Research, Institute for Advanced Studies, 1080 Vienna, Austria\\
$^{13}$ Department of Computational Neuroscience, University Medical Center Eppendorf, Hamburg University, Germany\\
$^{14}$ Department of Geography, University of Sheffield, S10 2TN, Sheffield, United Kingdom \\
$^{15}$ Department of Geography and Regional Research, University of Vienna, Universitätsstr. 7, A-1010 Vienna, Austria \\
$^{16}$ Department of Cultural Geography, University of Groningen, 9747 AD Groningen, Netherlands
}

\maketitle

\definecolor{shadecolor}{RGB}{180,180,180}

\begin{snugshade*}
\noindent{\textbf{Subject areas:} Complex Networks, Dynamical Systems, Computational Neuroscience 
\textbf{Author for correspondence:} Marc-Thorsten H\"utt\\
\textcolor{blue}{e-mail: m.huett@jacobs-university.de}
}
\end{snugshade*}

\newpage

\begin{abstract}
{\small The relationship of network structure and dynamics is one of most extensively investigated problems in the theory of complex systems of the last years. Understanding this relationship is of relevance to a range of disciplines -- from Neuroscience to Geomorphology. A major strategy of investigating this relationship is the quantitative comparison of a representation of network architecture (structural connectivity) with a (network) representation of the dynamics (functional connectivity). Analysing such SC/FC relationships has over the past years contributed substantially to our understanding of the functional role of network properties, such as modularity, hierarchical organization, hubs and cycles.

Here, we show that one can distinguish two classes of functional connectivity -- one based on simultaneous activity (co-activity) of nodes the other based on sequential activity of nodes. We delineate these two classes in different categories of dynamical processes -- excitations, regular and chaotic oscillators -- and provide examples for SC/FC correlations of both classes in each of these models. We expand the theoretical view of the SC/FC relationships, with conceptual instances of the SC and the two classes of FC for various application scenarios in Geomorphology, Freshwater Ecology, Systems Biology, Neuroscience and Social-Ecological Systems.

Seeing the organization of a dynamical processes in a network either as governed by co-activity or by sequential activity allows us to bring some order in the myriad of observations relating structure and function of complex networks.}

\end{abstract}

\section{Introduction}

  The relationship between network structure and dynamics has been at the forefront of investigation in the field of complex systems during the past decades, with networks serving as powerful abstract representations of real-world systems. However, a solid theoretical understanding of the generic features relating network structure and dynamics is still missing. Here, our strategy of investigating these features is via the quantitative comparison of network architecture ('structural connectivity', SC) with a network (or matrix) representation of the dynamics ('functional connectivity', FC). We establish key relationships using simple model representations of dynamics: excitable dynamics represented by a stochastic cellular automaton, coupled phase oscillators, chaotic oscillators represented by coupled logistic maps. We validate these relationships in coupled FitzHugh-Nagumo oscillators, in the excitable and the oscillatory regimes. Furthermore, we give examples of how the two classes of FC can be applied to various application domains, in which networks play a prominent role.

The simplest way of representing time series of dynamical elements as a network is to compute pairwise correlations. Often, one also knows about the 'true' or 'static' connectivity of these dynamical elements beforehand. The statistical question then arises in a natural way, whether the known network (SC) and the network derived from the dynamical observations (FC) are similar. As we will see in the applications, functional connectivity can either be thought of as dynamical similarities of nodes or flows (of material, activity, information, etc.) connecting two nodes. 

The simplicity of the dynamics included in our investigation allows us to work with this correlation-based approach. In case of a large heterogeneity of dynamical elements, very noisy dynamics, poor statistics (temporal sampling) or incomplete information, more sophisticated representations of dynamical relationships among nodes are required \cite{mukherjee2008network,marbach2010revealing,marbach2012wisdom,zhao2016part,newman2018network}.

Originating in Neuroscience \cite{honey2009predicting}, research into SC/FC correlations has become a promising marker for changes in systemic function and a means for exploring the principles underlying the relationship between network architecture and dynamics -- in Systems Biology \cite{sonnenschein2012network,ideker2012differential}, Social Sciences \cite{dong2018consensus,jalili2013social,potts2016exploring}, Geomorphology \cite{wainwright2011linking,baartman2020models, wohl2019connectivity} and Technology \cite{boguna2009navigability,Meyer2015,li2017complex}, just to name a few of the application areas.

Such SC/FC relationships are at the same time markers for certain forms of systemic behaviour (e.g., a loss of SC/FC correlation may indicate pathological brain activity patterns \cite{zhang2017disrupted}) and highly informative starting points for a mechanistic understanding of the system (e.g., revealing highly connected elements -- hubs -- as centres of self-organized excitation waves in scale-free graphs \cite{muller2008organization,moretti2020link}).

While the systemic implications and the key results have been reviewed elsewhere \cite{turnbull2018connectivity}, here we would like to show that across a range of dynamical processes and network architectures some fundamental common principles exist. We argue that one needs to distinguish between two types of functional connectivity, one related to synchronous activity (or 'co-activation'), the other related to chains of events (or 'sequential activation'). A system, like phase oscillators \cite{arenas2006synchronization,arenas2008synchronization,rodrigues2016kuramoto}, favouring one type of functional connectivity (for this example, synchronisation) can also display the other type of SC/FC correlations under certain conditions. 

A condition here is characterized by the network type, the strength of the coupling of the dynamical elements and the choice of further (intrinsic) parameters of each of the dynamical elements. Here we show many examples of transitions from one type of SC/FC correlations to another type under changes of these conditions. 

Stylized models of dynamics often offer a deep mechanistic understanding of the dynamical processes and phenomena and, in particular, help discern, how network architecture shapes the dynamical behaviour. The intense research over the past decades on networks of coupled phase oscillators as a stylized model of oscillatory dynamics illustrates this point -- with the topological determinants of synchronisability \cite{arenas2008synchronization,chen2009enhanced} and the lifetimes of intermediate synchronisation patterns in a time course towards full synchronisation and their relationships to the network's modular organization \cite{arenas2006synchronization} are two prominent examples of this line of investigation.

Remarkably, it is precisely this formal distinction between functional connectivity based on co-activation and sequential activation that is often hard to discriminate in more detailed (e.g., continuous) models \cite{messe2015closer} and experimental data \cite{haimovici2013brain}.

In the case of SC/FC correlations, the best investigated stylized model is the SER model of excitable dynamics \cite{muller2008organization,Garcia2012,fretter2017topological}. Key results include that the topological overlap \cite{messe2018toward} is highly associated with functional connectivity based in simultaneous activity, FC$_{\mbox{{\scriptsize sim}}}$, and that via this mechanism -- a clustering of high topological overlap values within modules -- modular graphs display high SC/FC correlations, while scale-free graphs tend to display low, or even systematically negative SC/FC correlations with this definition of FC \cite{Garcia2012,messe2018toward}. Furthermore, a large asymmetry of the sequential activation matrix (which is the foundation of functional connectivity based on sequential activation, FC$_{\mbox{{\scriptsize seq}}}$) can be associated with self-organized waves around hubs \cite{moretti2020link}. Additionally the role of cycles for organizing SC/FC correlations has been investigated \cite{Garcia2014} and in the deterministic limit of the model, a theoretical framework for predicting SC/FC correlations has been established \cite{messe2018toward}. 

\begin{figure}[!h]
	\centering
	\includegraphics[width= 0.9 \linewidth]{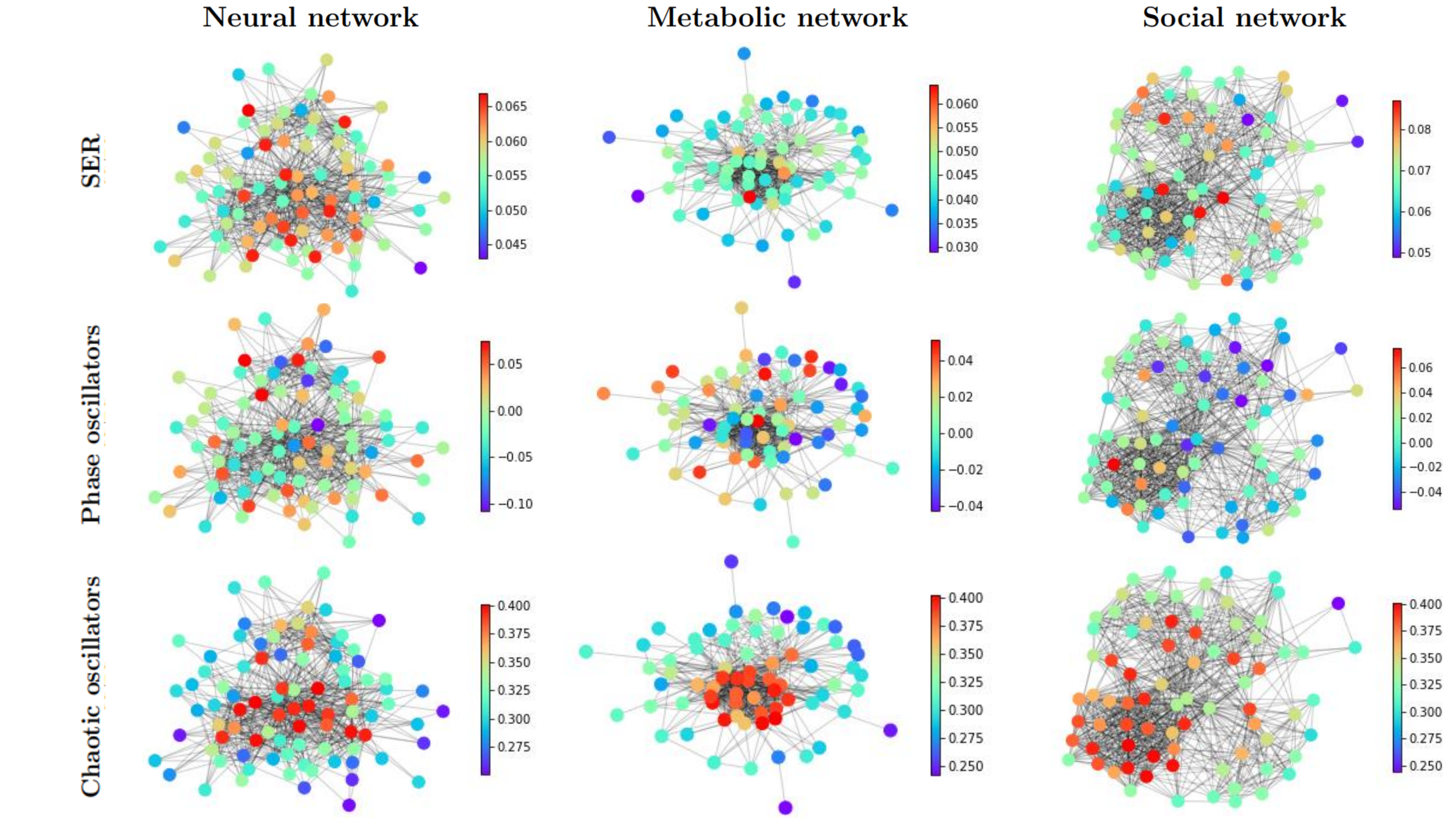}
	\caption{An illustrative example of applying different categories of dynamical processes to real networks with different structures. Neural: macaque cortical area network from \cite{markov2014weighted}. Metabolic: core metabolic system of the gut bacterium \textit{Escherichia coli} from \cite{palsson2015systems}. Social: skills awareness network from \cite{cross2004hidden}. SER model: The mean activity of each node after $1000$ timesteps, with a rate of spontaneous activity $f$=0.001 and a recovery probability $p$=0.1. Coupled phase oscillators: The average effective frequency of each node for ten simulations of length $T=200$ initialized with a uniform distribution of eigenfrequencies. Logistic map: The average standard deviation of the time series of each node for 10 simulations of 500 timesteps with the parameter $R$ for each node randomly selected from a uniform distribution with $R_{\mbox{{\scriptsize min}}} = 3.7$ and $R_{max} = 3.9$.}
	\label{fig0}
\end{figure}

As a first illustration of the tremendous power of probing networks with various types of dynamics, in order to understand how network architecture determines some of the dynamical features, in Figure \ref{fig0} we show snapshots of dynamical states for three real-life networks coming from different domains -- neuroscience (the macaque cortical area network from \cite{markov2014weighted}), systems biology (the core metabolic system of the gut bacterium \textit{Escherichia coli} from \cite{palsson2015systems}) and social sciences (intra-organizational network of skills awareness in a company from \cite{cross2004hidden}) -- under the action of three types of dynamics -- excitable dynamics, phase oscillators, the logistic map as an example of a chaotic oscillator.

The real-life networks shown in Figure \ref{fig0} can all be considered examples of structural connectivity. The detailed description of the structure of these networks is given in the Supporting Information.

An important question around Figure \ref{fig0} is, whether the three types of dynamics are plausible for the networks at hand. First, we would like to emphasize that the strategy of our investigation is to probe network architectures by simple prototypes (or very stylized forms) of dynamics, rather than devising realistic models of the most plausible form of dynamics for each of these networks. 

In the case of the cortical area network the excitable dynamics, as well as the oscillatory dynamics can be seen as stylized but plausible dynamical probes and, in fact, those have been previously employed to explore such network structures \cite{wu1999selective, moretti2020link, corchs2001neurodynamical, messe2014relating}. But also chaotic dynamics, as in the third row of Figure \ref{fig0}, have been used to study neuronal connectivity patterns \cite{babloyantz1994computation,xu2018synchronization}. 

In the case of metabolic networks, synchronous activity patterns, and hence coupled phase oscillators, are a plausible form of dynamics (see, for example, the arguments in \cite{becker2011flow}, where enzymes are described as cyclically operating devices, as well as the prominent usage of correlation networks in metabolomics \cite{camacho2005origin,muller2007consistency,kumar2020correlation}).
A more pathway-oriented view of metabolism might emphasize the propagation of activity and, hence, would be closer to the excitable dynamics shown in the first row of Figure \ref{fig0}. Chaotic oscillators are clearly less relevant for this application domain. 

Interaction dynamics, contact dynamics and information flow in a corporate setting unite aspects of excitable dynamics (as in the case of rumour spreading, \cite{moreno2004dynamics}) or synchronisation \cite{helbing2012social,nowak2020synchronization}. But also chaotic dynamics have been employed to model decision dynamics and activity in corporate settings \cite{dooley1999explaining,whitby2001non,yuan2019understanding}.

The three main messages of the illustration of dynamics on real-life networks shown in Figure \ref{fig0} are: (1) The representation of complex systems as networks enables the probing of such complex structures with dynamics. (2) Different networks react differently to one type of dynamics. This general point can be seen for example in Figure \ref{fig0} by following one type of dynamics (e.g., excitable dynamics; first row in Figure \ref{fig0}) across the three networks and observing that groups of nodes acting together (similar colour, representing similar dynamical states) can be either in the periphery or in the centre in this spring-embedding representations of these networks. (3) A given network reacts differently to different dynamical probes. This general feature can be seen by following a single network across different types of dynamics (a column in Figure \ref{fig0}). Regions in the graph with a similar dynamical state (same colours) for one dynamics look heterogeneous (different colours) for another dynamics.  Also, similarities occur. The periphery and the centre of the networks tend to behave differently in all the examples of dynamics shown in Figure \ref{fig0}. 

It is obvious that such an illustration can only provide a single snapshot of the diverse dynamics possible on such networks, even for a single type of dynamics, as the internal parameters at each node, as well as the coupling type and strength among them can have different values. In the following, we want to further explore the systematic changes of these dynamical patterns as a function of network architecture, coupling and internal dynamical parameters and how this theoretical framework can be applied to various disciplines.

\section{Results and Discussion}

We create different instances that indicate the behaviour of the two classes of FC using various numerical schemes. The means of enhancing or destroying SC/FC correlations can be structural (i.e., driven by network architecture) or dynamical (induced by changing the parameters of the dynamical model). The investigation is organized around the form of change: (a) topological changes, (b) changes in coupling strength, (c) changes of the intrinsic parameters of the individual elements. In Section (d) we illustrate these principles further in a case study on a network of coupled FitzHugh-Nagumo oscillators undergoing a transition from excitable to oscillatory behaviour. 
Using the three examples from Figure \ref{fig0}, in Section (e), we show the behaviour of SC/FC correlations on these real-world networks.

\subsection{Topological changes}
The first part of our investigation is related to the effect of topology in the SC/FC correlations. We started with networks with a distinct structure (modular graph, hierarchical graph, regular graph), which we gradually destroyed either by randomizing or by rewiring the initial network (see Methods).
\begin{figure}[!h]
	\centering
	\includegraphics[width=  \linewidth]{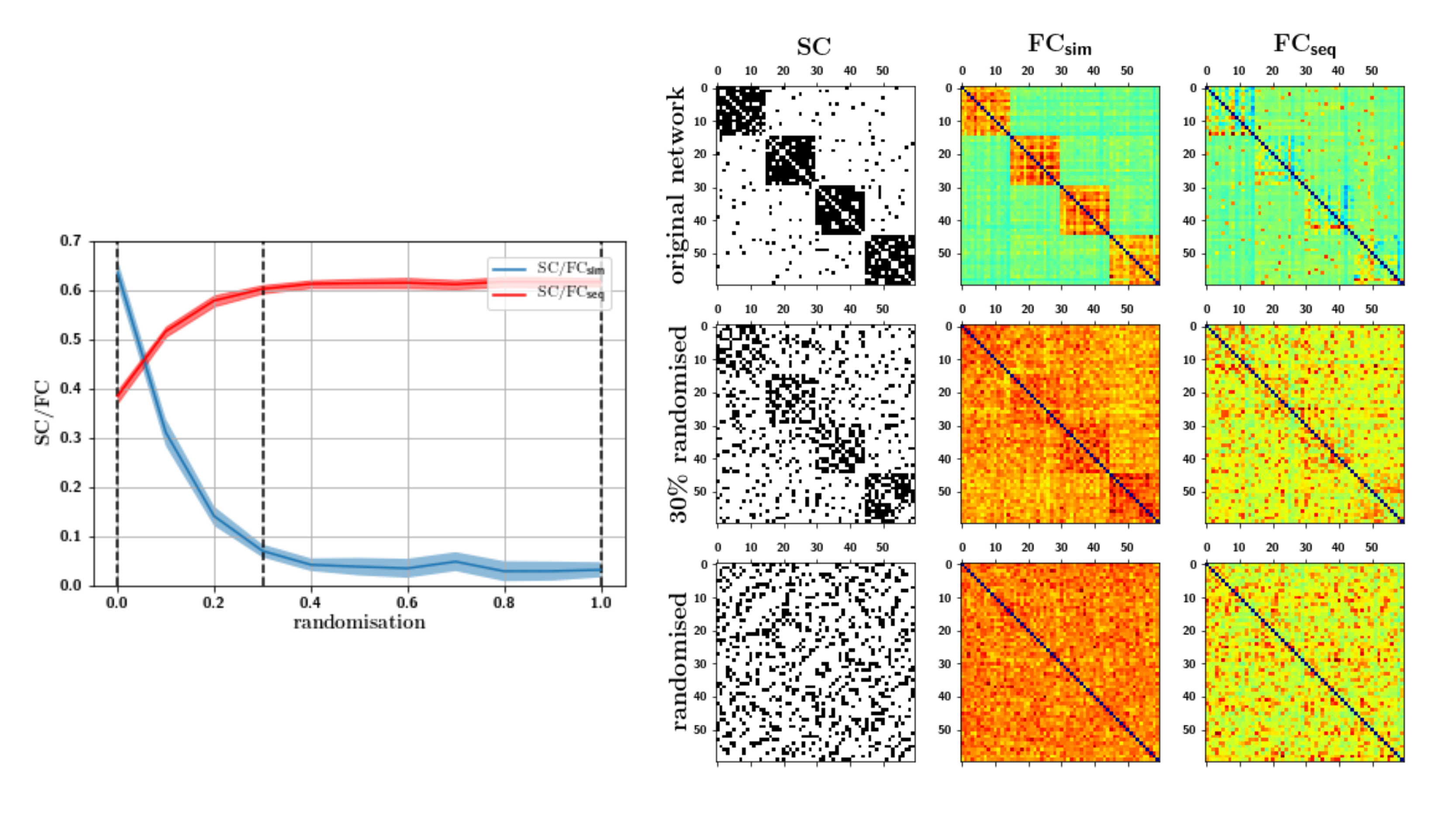}
	\caption{\textbf{Left}: SC/FC$_{\mbox{{\scriptsize sim}}}$ and SC/FC$_{\mbox{{\scriptsize seq}}}$ correlations across the randomisation of a modular network. \textbf{Right}: Illustration of the SC and the FC$_{\mbox{{\scriptsize sim}}}$, FC$_{\mbox{{\scriptsize seq}}}$ matrices for three network cases, pointed out by the dashed vertical black lines on the left figure (original modular network, 30\% randomised network and completely randomised network). The dynamical model used for the FC is the SER model (parameters: $t_{max}$=10, $N_R$=10000, $p$=0.1, $f$=0.001.)}
	\label{fig2}
\end{figure}

Figure \ref{fig2} introduces the comparison of structural connectivity and functional connectivity on the matrix level, by depicting the adjacency matrices of two networks, together with examples of the corresponding functional connectivity matrices derived from dynamics (here: the co-activation and sequential activation matrices obtained from simulations of the SER model; see Methods). This matrix view on SC/FC relationships is similar to Figure 1 in \cite{messe2015closer} and Figure 1 in \cite{Garcia2012} and allows us to visually discern the strong positive correlation between the adjacency matrix and the co-activation matrix in the case of the modular graph (first row) and the apparent lack thereof in the more random graph (second row), for which we, however, can visually perceive an agreement between the adjacency matrix and the sequential activation matrix. So, here a change in network topology goes along with a change from one type of SC/FC correlations (co-activation to sequential activation). This is the phenomenon we set out to explore further in the following.

Figures \ref{figS1} and \ref{figS2} shows the same matrix view, but for coupled phase oscillators and logistic maps, respectively. In Figure \ref{figS1} (phase oscillators), visual inspection clearly shows that the SC/FC correlations based on sequential activation is much weaker than the one based on co-activation. Also, SC/FC$_{\mbox{{\scriptsize sim}}}$ remains visibly high during randomisation. In Figure \ref{figS3} (logistic maps), the lack of correlation between co-activation and the modular structure is clearly seen, as is the (faint, but discernible) agreement of this modular structure with sequential activation. Careful visual inspection also reveals the persisting positive SC/FC$_{\mbox{{\scriptsize seq}}}$ correlation, as well as the negative SC/FC$_{\mbox{{\scriptsize sim}}}$ correlations, under randomisation of the modular network. Figure \ref{figS3} shows examples of space-time plots for single runs of the chaotic dynamics and thus provides a microscopic view of the results summarized in Figure \ref{fig2}.

\begin{figure}[!h]
	\centering
	\includegraphics[width=12.5cm]{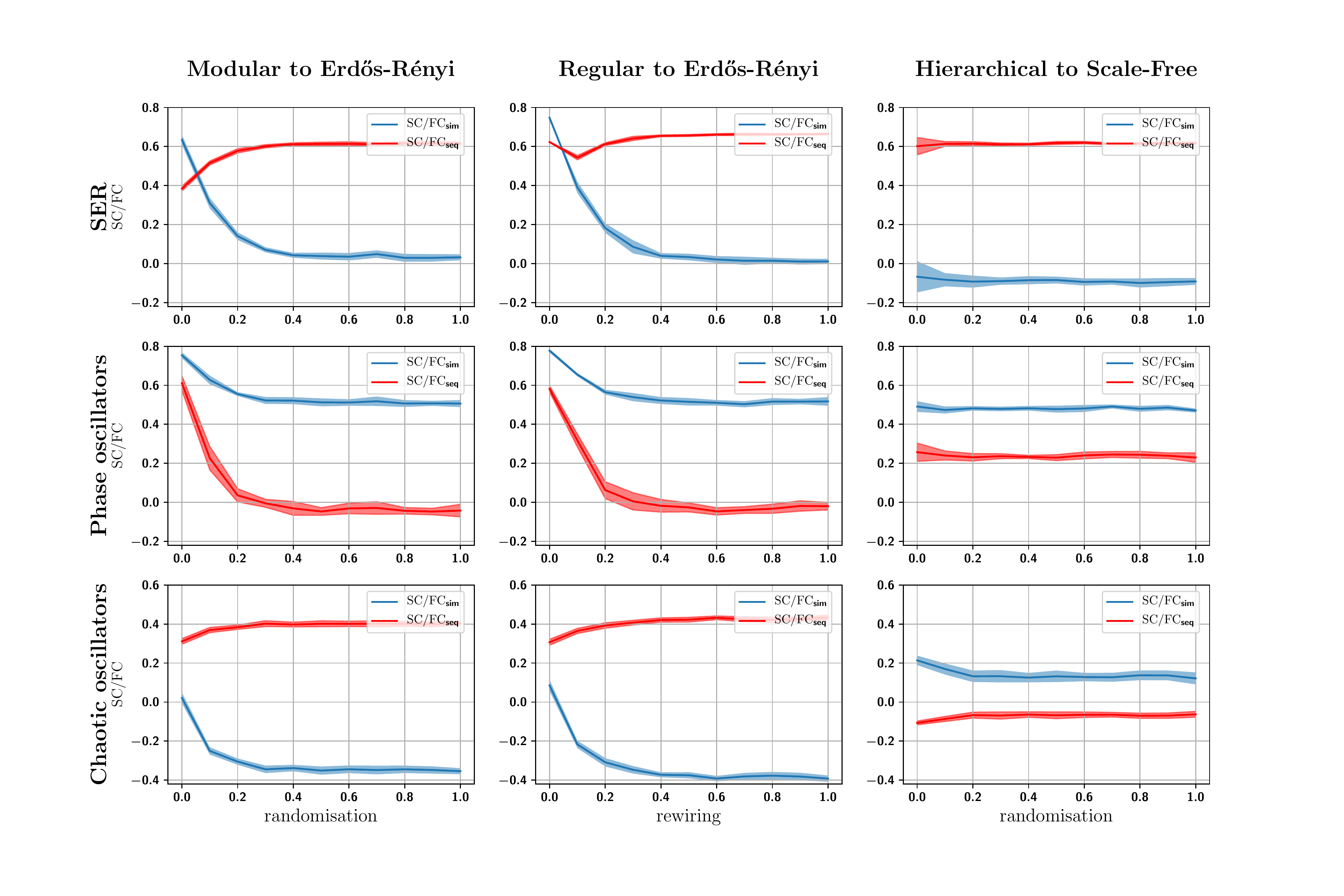}
	\caption{SC/FC$_{\mbox{{\scriptsize sim}}}$ and SC/FC$_{\mbox{{\scriptsize seq}}}$ correlations across the range of randomisation/rewiring processes.
 \textbf{First column}: randomisation of a modular graph.
\textbf{Second column}: rewiring of a regular graph.
\textbf{Third column}: randomisation of a hierarchical graph in the three models.
\textbf{First row}: SER model; parameters: $t_{max}$=10, $N_R$ (over different initial conditions)=10000, $N_R$ (over different initial graphs)=10, $p$=0.1, $f$=0.001.
\textbf{Second row}: Coupled phase oscillators; parameters: $t_{max}$=50, $N_R$ (over different initial conditions)=100, $N_R$ (over different initial graphs)=10, $\omega \in$ (0,1), $k$=10, $\sigma$=0.25, $u\in$ (0,1).
\textbf{Third row}: Logistic map (chaotic oscillators); parameters: $t_{max}$=500, $N_R$ (over different initial conditions)=50, $R\in$ (3.7, 3.9), $k$=2.}
	\label{fig3}
\end{figure}

In Figure \ref{fig3}, we go from rather structured network topologies to rather unstructured random network topologies.  Figure \ref{fig3} supports the visual impression from the matrix examples shown in Figure \ref{fig2} by showing the two types of SC/FC correlations as a function of network randomisation procedures, for the SER model (which was also used in Figure \ref{fig2}), as well as two other types of dynamics, namely coupled phase oscillators and coupled logistic maps in the chaotic regime (see Methods). It should be noted that each of these dynamical models has been instrumental in the past in advancing our understanding of fundamental relationships between network architecture and dynamics (see, e.g., \cite{muller2008organization,messe2018toward,damicelli2019topological} for the SER model, \cite{arenas2006synchronization,rodrigues2016kuramoto} for coupled phase oscillators, and \cite{lind2004coherence,masoller2011complex} for the logistic maps).

For the SER model, we see a trend that structured topologies favour high SC/FC correlations of both types, whereas unstructured random networks favour high SC/FC$_{\mbox{{\scriptsize seq}}}$ correlations. We can also see that SC/FC$_{\mbox{{\scriptsize sim}}}$ is very sensitive to topological changes, in contrast to SC/FC$_{\mbox{{\scriptsize seq}}}$, which, in this case, shows a more stable behaviour. The networks of coupled phase oscillators behave in almost the opposite way, where co-activation (rather than sequential activation) is favoured by random network structures and shows a more stable behaviour under topological changes.
In the case of chaotic oscillators, the details about the network architecture and the selection of the type of coupling matter. For this case the transition from structured to unstructured networks does not affect the SC/FC$_{\mbox{{\scriptsize seq}}}$, but leads to strong negative correlations of SC/FC$_{\mbox{{\scriptsize sim}}}$. The hierarchical network is the only one, though, in which the destruction of the modularity is not revealed from the dynamics. For this graph, all the dynamical models show that the randomisation does not affect qualitatively the value of SC/FC correlations, instead constant, low positive correlations of SC/FC$_{\mbox{{\scriptsize seq}}}$ and constant, low negative correlations of SC/FC$_{\mbox{{\scriptsize sim}}}$ are maintained during the randomisation process.

\subsection{Changes in coupling strength}

The second set of our numerical experiments pertains to changes in the coupling strength among nodes. For this type of change, only the models of the phase and chaotic oscillators can be used, as the SER model in the form used here has no coupling parameter (which could, however, be introduced via a relative excitation threshold, as in \cite{hutt2012stochastic,fretter2017topological}). 

For phase oscillators all the network architectures stabilise SC/FC$_{\mbox{{\scriptsize sim}}}$ against changes of coupling strength. Large values of coupling strength lead to rapid synchronisation (co-activity of the nodes) and therefore to inadequate amount of information for the sequential activation. As a result, seeing the structure of the network through the dynamics using the sequential activation is, in this case, not possible.
For the chaotic oscillators we observe general trends of increasing SC/FC$_{\mbox{{\scriptsize seq}}}$ with increasing coupling, reaching a maximum, and gradually decreasing for further increase of the coupling, essentially across all network architectures.

\begin{figure}[!h]
	\centering
	\includegraphics[width=12.5cm]{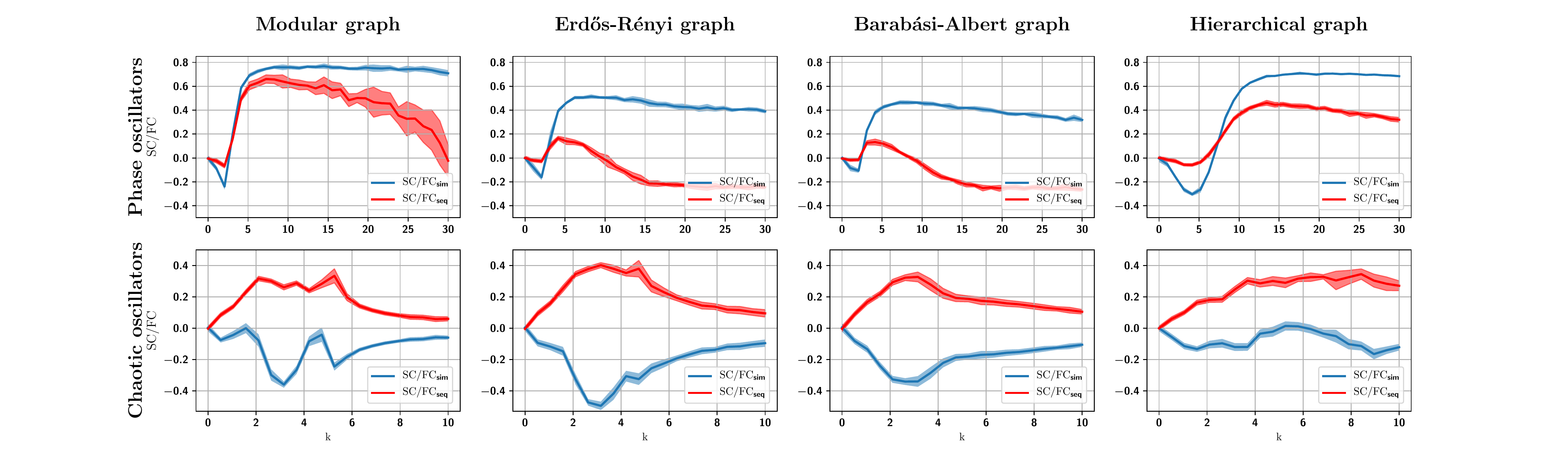}
	\caption{SC/FC$_{\mbox{{\scriptsize sim}}}$ and SC/FC$_{\mbox{{\scriptsize seq}}}$ correlations as a function of the coupling strength among the nodes in the two types of oscillators applied to different network architectures. \textbf{First column}: Modular graph. \textbf{Second column}: Erd\H{o}s-R\'enyi graph. \textbf{Third column}: Barab\'{a}si-Albert graph. \textbf{Fourth column}: Hierarchical graph. \textbf{First row}: Coupled phase oscillators; parameters: $t_{\mbox{{\scriptsize max}}}$=50, $N_{\mbox{{\scriptsize R}}}$ (over different initial conditions)=100,  $N_{\mbox{{\scriptsize R}}}$ (over different initial graphs)=10, $\omega\in$ (0,1), $k$=10, $\sigma$=0.25, u $\in$ (0,1). \textbf{Second row}: Logistic map (chaotic oscillators); parameters: $t_{\mbox{{\scriptsize max}}}$=500, $N_{\mbox{{\scriptsize R}}}$=50, $R\in$ (3.7, 3.9), $k$=2.}

	\label{fig4}
\end{figure}

\subsection{Changes in intrinsic parameters}
Each dynamical model is characterized by specific intrinsic parameters that determine the behaviour of the individual elements and of the system, too. Changes in the values of the intrinsic parameters may result in drastic changes to the functional connectivity. In this part of the investigation, the two types of the functional connectivity are studied as a function of such intrinsic parameters. We are here attempting to address the following question:
Is there at least one class of the functional connectivity that can survive under the changes of a dynamical parameter of the model? Or relatedly, is it possible to observe the structure of the network through the dynamics even if we are consistently changing an intrinsic parameter?

The stochastic SER model is characterized by the recovery probability, $p$, that determines if a node in the refractory state will return to the susceptible state. For the phase oscillators we use the range of natural frequencies as the intrinsic parameter. 
The logistic map has only one intrinsic parameter, $R$, which defines the dynamical behaviour of the uncoupled oscillator. We here vary the average $R$ such that the uncoupled oscillator would reside in the chaotic regime (3.56, 4.0).

\begin{figure}[!h]
	\centering
	\includegraphics[width=12.5cm]{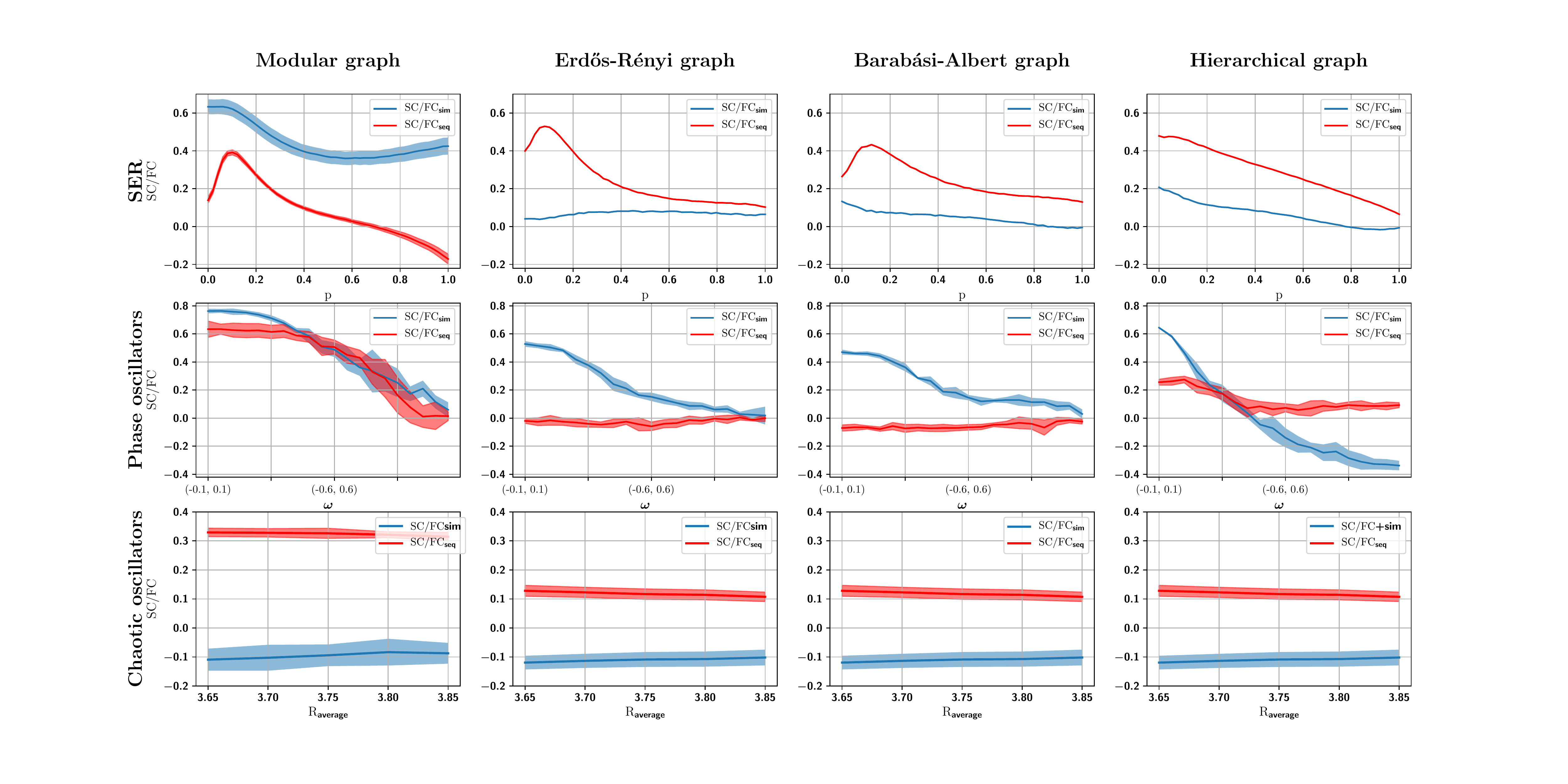}
	\caption{SC/FC$_{\mbox{{\scriptsize sim}}}$ and SC/FC$_{\mbox{{\scriptsize seq}}}$ correlations under changes of dynamical parameters in the three models.
\textbf{First row}: SER model with increasing the recovery probability (parameters: $t_{max}$ =10, $N_{R}$=10000, $f$=0.001).
\textbf{Second row}: coupled phase oscillators  under a widening of the distribution of the natural frequencies (parameters: $t_{max}$=50, $N_{R}$=100, $k$=10, $\sigma$=0.25, $u\in$ (0,1)).
\textbf{Third row}: logistic map  under a shift of $R_{average}$  of the distribution of $R$ within the interval (3.6, 3.9), keeping the width equal to 0.2 (parameters: $t_{max}$=500, $N_{R}$=50, $k$=2). Four network architectures were used for each model: \textbf{First column}: modular graph. \textbf{Second column}: Erd\H{o}s-R\'enyi graph. \textbf{Third column}: Barab\'{a}si-Albert graph. \textbf{Fourth column}: hierarchical graph.}
	\label{fig5}
\end{figure}

Figure \ref{fig5} shows the results of this part of the investigation. For the SER model we see that network effects are consistent across the whole parameter range. We can see that SC/FC$_{\mbox{{\scriptsize sim}}}$ is consistently high for the modular graph and very close to zero for all the other graphs, where, in contrast, the SC/FC$_{\mbox{{\scriptsize seq}}}$ has positive correlation values. For the phase oscillators, the width of the frequency distribution matters: increasing width leads to a consistent decrease of SC/FC$_{\mbox{{\scriptsize sim}}}$, but leaves SC/FC$_{\mbox{{\scriptsize seq}}}$ intact in all graphs, except for the modular, in which the behaviour of SC/FC$_{\mbox{{\scriptsize seq}}}$ is similar to SC/FC$_{\mbox{{\scriptsize sim}}}$. The logistic map does not show any parameter sensitivity of SC/FC correlations in the different network architectures.

\subsection{Additional case study}

\begin{figure}[h!]
	\centering
	\includegraphics[width= 0.85\linewidth]{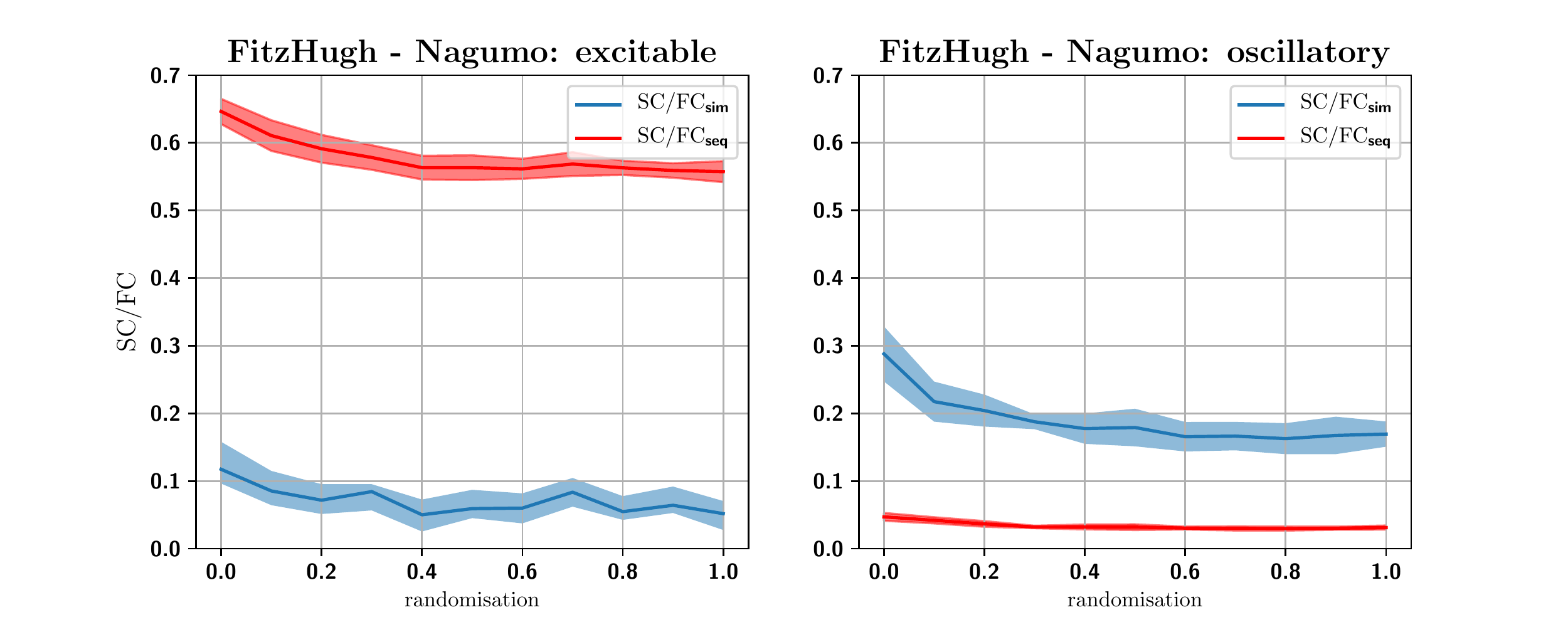}
	\caption{SC/FC correlations from the FitzHugh-Nagumo model in excitable (left) and oscillatory (right) regime while randomizing random modular networks. The blue curves represent co-activation (i.e., a time window of 1 ms), while the red curves represent sequential activation (i.e. using a time window of 12 ms).}
	\label{fig6}
\end{figure}

The FitzHugh-Nagumo model can be used for a case study verifying whether our previous results translate to this more detailed, more realistic model. To this end, we study the behaviour of SC/FC correlations as a function of randomizing a modular graph in the oscillatory regime ($a$=0) and in the excitable regime ($a$=0.8). The results of this more complicated model shown in Figure \ref{fig6} confirm the general observations derived from the two corresponding minimal models: The excitable dynamics enhance the SC/FC$_{\mbox{{\scriptsize seq}}}$ across the transition of a modular to an Erd\H{o}s-R\'enyi graph, whereas oscillations favours the SC/FC$_{\mbox{{\scriptsize sim}}}$ across the randomisation process.

\subsection{SC/FC correlations in real networks}

We can now return to the real-life networks from Figure \ref{fig0} and study the two types of SC/FC correlations in these networks as a function of the intrinsic parameters of the dynamical models, as done in Figure \ref{fig5} for the abstract network architectures. 
\begin{figure}[!h]
	\centering
	\includegraphics[width=12.5cm]{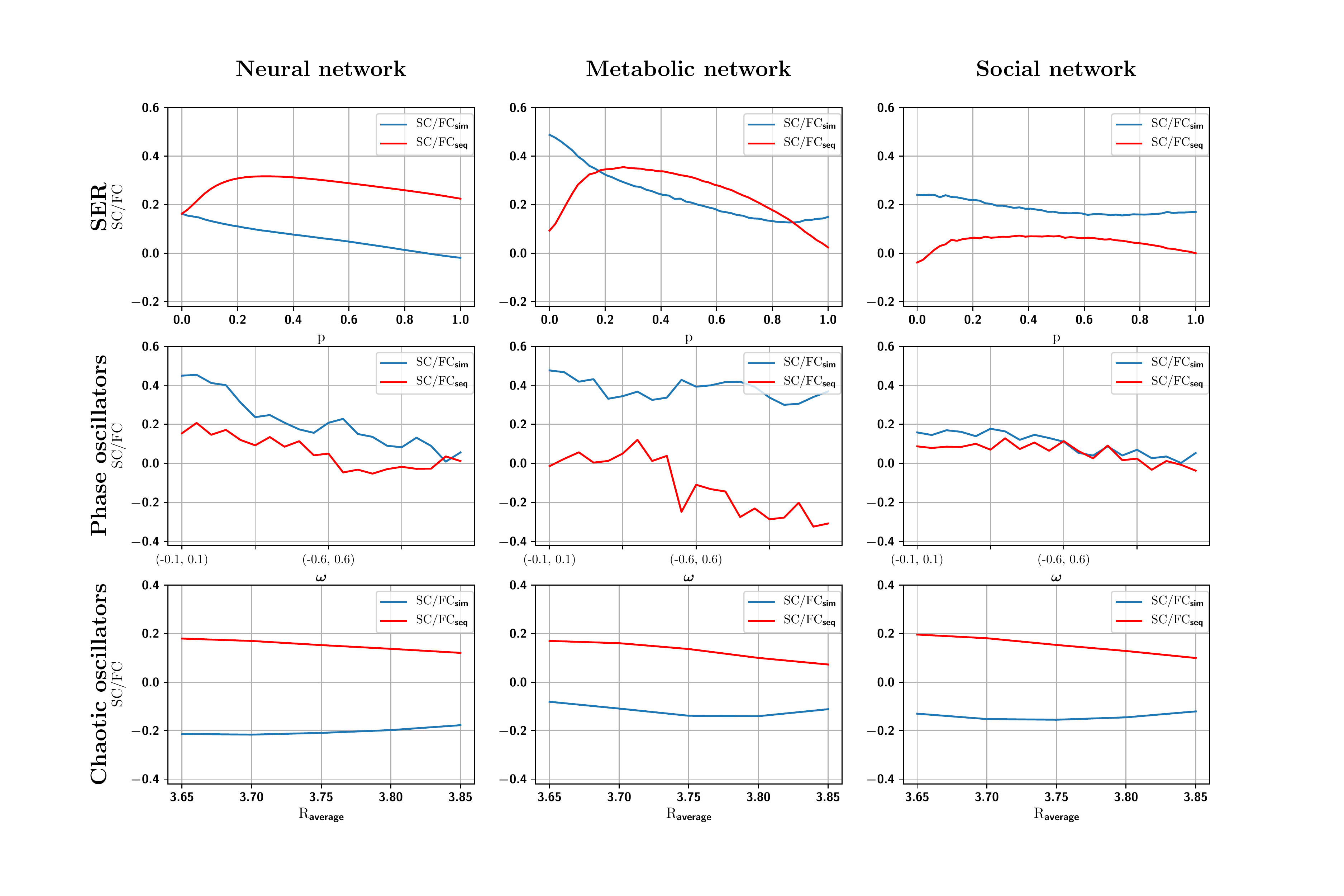}
	\caption{SC/FC$_{\mbox{{\scriptsize sim}}}$ and SC/FC$_{\mbox{{\scriptsize seq}}}$ correlations for the three real-life networks under changes of dynamical parameters in the three models, as in the section (c). \textbf{First column:} neural network. \textbf{Second column:} metabolic network. \textbf{Third column:} social network. \textbf{First row:} SER model; parameters: $t_{max}$=10, $N_R$ (over different initial conditions)=10000, $f=0.001$. \textbf{Second row:} coupled phase oscillators; parameters: $t_{max}$=50, $N_R$=100 (over different initial conditions), $k$=10, $\sigma$=0.25, $u\in$ (0,1). \textbf{Third row:} logistic map (chaotic oscillators); parameters:  $t_{max}$=500, $N_R$=50, $k$=2.}
	\label{fig7}
\end{figure}
The results are summarised in Figure \ref{fig7}. Regarding the SER model, we see high SC/FC$_{\mbox{{\scriptsize seq}}}$ correlations for the neural system and, in contrast, high SC/FC$_{\mbox{{\scriptsize sim}}}$ correlations for the social system, under the increase of the recovery probability, while in the case of the metabolic system, it depends on the parameter value, which of the two types of SC/FC correlation is higher. For the phase oscillators, we see initially high correlations that approach zero value, as we increase the width of the eigenfrequencies distribution, with the SC/FC$_{\mbox{{\scriptsize sim}}}$ to have constantly higher values. In the metabolic network, dominant and relatively strong and stable SC/FC$_{\mbox{{\scriptsize sim}}}$ appears under the same changes of the $\omega$ distribution, whereas the zero values of SC/FC$_{\mbox{{\scriptsize sim}}}$ for the narrow distributions give place to strong negative correlations as we move to wider distributions. The behaviour of the social network is similar to the neural one, but with lower SC/FC correlation values. The results for the logistic map are dominated by SC/FC$_{\mbox{{\scriptsize seq}}}$ correlations, independent of network architecture and parameter value.

\section{Applications}

In this section we briefly review some areas of application to illustrate, (1) how structural connectivity can be defined in these contexts, (2) which approaches for defining functional connectivity exist in this domain, and (3) how the two types of functional connectivity appear in this setting. 

Throughout this investigation, we have the following scenario in mind. Given a network (structural connectivity) and dynamical processes for each of the nodes, we analyse the time series observed at each node and derive relationships among the nodes (functional connectivity), in order to understand how network architecture determines or shapes the dynamical relationships among nodes. This interplay of structure and dynamics is then illustrated by and quantified in terms of SC/FC correlations.  
This clear distinction between (static or slowly changing) structural connectivity, which serves as 'infrastructure' for dynamical processes, and (often rapidly changing) functional connectivity is not plausible for all applications. As a consequence, a debate about SC/FC correlations is not possible in important areas of research. Often, in those disciplines the evolution of the network itself under the action of its agents (nodes) is investigated. Social network analysis (SNA) is the methodology of choice for such situations \cite{prell2012social} (see Supporting Information for more details).

\subsection{Application to Geomorphology}

Within hydrology and geomorphology, the examples of structurally connected pathways that we will discuss here are those that direct the flow of water and sediment over the surface and within the near-surface zone. On steeper slopes, these structurally connected pathways are predominantly controlled by the topography and vegetation, whereas on slopes (c. $<5^\circ $) other surface characteristics such as microtopography and soil hydraulic properties can become relatively important. 
The dynamical processes occurring over this structural template and subsequent functional connectivity are then an emergent property of these structural controls in combination with dynamical inputs (e.g. precipitation).
The presence of vegetation also (i) modifies soil properties, (ii) often has an associated microtopography and (iii) can impede/reduce flows due to friction and damming effects, so that there are dynamic feedbacks between the structural and functional connectivity \cite{wainwright2011linking,connecteur2018better,baartman2020models}.

There are various approaches to assessing structural connectivity in hydrology and geomorphology. If we take a river network, the structural connectivity of the network can be defined based on the pathways connecting all links through which water can potentially flow, resulting in a graph structure most often in the form of a tree \cite{Eros2012,schick2007directed}, with the exception of braided streams \cite{marra2014network}, deltas \cite{tejedor2015delta,passalacqua2017delta} or, for a more broad example, coastal sediment pathways \cite{pearson2020sediment}. Thus, the structural connectivity of river networks can be quantified using, for example, the pairwise connectivity of its underlying tree structure – an approach that has been used both for natural and synthetic river networks \cite{trigg2013surface}. These synthetic networks are useful as they inform our mechanistic understanding of these complex systems. One such example is optimal channel networks (OCNs) which can be generated for varying values of the energy exponent (a parameter that characterizes the mechanics of erosion processes in channel formation) in order to reveal how such topological factors, lead to emergent network properties \cite{abed2017emergent}. OCNs replicate the major scaling features associated with river networks around the world \cite{rigon1993optimal,balister2018river}, and thus bridge the gap between random graphs shown in Figure 3 that go from structured to unstructured network topologies and river networks observed in nature. Furthermore, river networks have a directional structural template, with links connecting high-elevation nodes to low-elevation nodes. On hillslopes, structural connectivity has been measured based on the upslope contributing area to a particular node (e.g. \cite{jencso2009hydrologic}), and on the combination of topographically connected flow paths (using flow routing algorithms) and the presence of vegetation (measured using remote sensing techniques) that intersects (and in certain environments disconnects) these flow paths (e.g. \cite{turnbull2019structure}). Similarly, in hydrological analysis of sub-surface flow, structural connectivity of a network of wells may be determined from the downslope direction of surface topography from any well (e.g. \cite{rinderer2018assessing}).

In these examples from hydrology and geomorphology, we are concerned with (1) areas that have a similar response to a dynamical probe (e.g. rainfall event); and (2) connectivity of fluxes, i.e. flows of water and/or sediment that are transported through the network to a downslope/stream location. These two types of functional connectivity map onto FC$_{\mbox{{\scriptsize sim}}}$ and FC$_{\mbox{{\scriptsize seq}}}$ respectively. Approaches used to measure FC in hydrology and geomorphology are varied.

In relation to (1), FC$_{\mbox{{\scriptsize sim}}}$ geostatistical analysis is often used to assess how the scales of co-activation change in response to a dynamical probe. For example, one can measure the autocorrelation of soil-moisture content and how this changes over time, both in response to a rainfall event and then during a refractory period (see for example \cite{turnbull2010changes}). In the case of sub-surface flows, FC between two wells (nodes) a and b is deemed to exist if well b is downslope of well a and the wells are co-activated (i.e. water is present in both) (e.g. \cite{phillips2011connectivity,zuecco2019quantification}). 

In relation to (2), FC$_{\mbox{{\scriptsize seq}}}$ is often assessed/inferred based on gauges within a network being activated at a range of lag times, thus indicating the flow of water or sediment between the two locations \cite{keefer2008long}. Geostatistical analysis has also been used to quantify how FC$_{\mbox{{\scriptsize seq}}}$ changes throughout a flood event \cite{trigg2013surface}.
The FC$_{\mbox{{\scriptsize seq}}}$ of fluxes through a real or synthetic river network has also been simulated and incorporated into a dynamic tree approach by analyzing dye propagation models at successive snapshots \cite{zaliapin2010transport}. Such approaches could be particularly valuable for studying the little understood impact of pulses of sediment \cite{czuba2014network}, nutrients \cite{sanchez2010changes} or other diffuse chemicals \cite{ashauer2007modeling} transported by surface waters. In the case of sub-surface flows, sequential activity of the two wells may be inferred from time-series analysis of water levels at a range of time lags (e.g. \cite{rinderer2018assessing}). 

Suitable empirical data for measuring these examples of functional connectivity are relatively scarce, and therefore researchers often turn to process-based modelling as a way to quantify both types of functional connectivity. For example, high spatio-temporal resolution modelling can be used to measure times during a storm event when infiltration will be locally satisfied and thus the onset of runoff generation (excited) or not (susceptible/refractory) due to spatial variability in infiltration capacity, rainfall intensity and antecedent soil-moisture content. From this high spatio-temporal modelling, the degree of synchronised functional connectivity of all locations within the model spatial domain can be derived. The spatial pattern of these synchronised points in turn determines the sequential connectivity of runoff and sediment flux \cite{parsons1997distributed, mueller2007impact}. For example, using high-resolution process-based modelling \cite{turnbull2018connectivity} measured the length of connected flow paths on grass and shrub hillslopes that had varying lengths of structurally connected pathways. In this example, the longer the SC, the higher the FC$_{\mbox{{\scriptsize seq}}}$ of discharge and sediment flux, which is similar to that observed in the case of coupled phase oscillators where FC$_{\mbox{{\scriptsize seq}}}$ is destroyed with an increase in the randomness of the network. 

Whereas some evidence exists for the impact of SC on FC$_{\mbox{{\scriptsize seq}}}$ at a particular timescale \cite{turnbull2018connectivity}, there remains scope to examine patterns of FC$_{\mbox{{\scriptsize seq}}}$ both at increasing lag times and in response to dynamical probes of different magnitudes for their impact on landscape change (topological changes). Furthermore, geomorphological assessment of the importance of coupling strength among nodes remains unexplored.
An important point to highlight is that timescales of synchronised versus sequential functional connectivity in hydrology and geomorphology are often markedly different. The widespread synchronisation of activity over a spatial range is valid for a small time period (mins/hours) whilst the sequential propagation of fluxes through the network occurs over longer time periods – hours to days to decades – depending on the size/configuration/connectivity of the network/system. Similarly, earthquake/storm-driven landslides tend to be synchronised over timescales of hours-days, whereas the resulting cascade of material through the network are sequential over significantly longer timescales (e.g., \cite{korup2005large,croissant2019seismic,tunnicliffe2018reaction}). 

Likewise, the spatial scales associated with synchronised and sequential connectivity tend to differ. For example, nearby nodes often exhibit synchronisation, whereas sequential flux propagation is observed at a larger spatial scale \cite{betterle2017characterizing}. Flood events highlight the potential for sequential propagation of processes over large spatial scales over time periods of hours to days. In catastrophic flooding in the Lockyer valley in Queensland, Australia in 2011 the hydrological and sedimentological connectivity between the channel and the floodplain was spatially variable depending on the morphology of the reach and whether it was expanding or contracting \cite{croke2013channel}. Hence, in this example, the organisation of dynamical processes in the network was crucial to the change in channel morphology, despite assumptions that in such a large flooding event thresholds for connectivity would have been exceeded.

\subsection{Application to Freshwater Ecology}

Over the decades, ecosystem ecology has developed a considerable amount of methodologies for network analysis, which contributed to the characterization of the evolution and status of ecosystems \cite{borrett2019walk}. The structural connectivity (SC) is represented in these models by depicting standing stocks (e.g., biomass, local communities or populations, species, individuals, or habitat patches) as nodes, and the interactions between them (e.g., feeding, the movement of animals or diseases) as links \cite{borrett2019walk}. Within landscape connectivity, the spatial structure of river networks (SC) plays a key role in structuring ecological patterns \cite{siqueira2014predictive}. Graph representations of river networks are often modelled to resemble the hierarchical structuring of habitat patches (nodes) and the potential dispersal corridors (links) \cite{bodin2010ranking, erHos2012characterizing, erHos2019landscape, minor2008graph, bohlen2001plant}. 

Dynamical approaches have not explicitly used the terms co-activation or sequential activation for describing functional connectivity. However, some of the notions in this paper can also be deduced from dynamical approaches of habitat connectivity already applied in aquatic ecology. The focus on animal movement and dispersal has been driving the theoretical and empirical work in the past few decades \cite{baldan2020multi, erHos2019landscape, heino2017integrating, saura2010common, schick2008understanding, schick2007directed, tonkin2018role}, especially in the light of fragmented landscapes. In models of organismal-environment interactions based on landscape’s resistance to dispersal \cite{calabrese2004comparison, neufeld2018incorporating} and in models that include the intrinsic dispersal abilities and limitations of organisms (i.e., individual-based population or metacommunity models \cite{larsen2021geography, muneepeerakul2007neutral, siqueira2014predictive}) the movement of animals is represented as the dispersal of individuals from node to node (an analogous of the flow of vehicles in a transport network \cite{heino2017integrating}.

Community ecologist have long seen individual populations and communities as oscillators \cite{gilarranz2015inferring}. They focused on the dynamics of a modular network, inferred from the synchrony between the rate of change of the population density within nodes \cite{gilarranz2015inferring}. Another example is the ecohydrological study of \cite{larsen2021geography}, where they proposed the concept of “fluvial synchrograms” to explain patterns of the geography of metapopulations synchrony within a river network, using the case of freshwater fishes of Europe. In their empirically driven approach based on the geography of synchrony, they developed theoretical synchrograms using simulated time-series of species abundance from the spatially explicit dynamic metacommunity model \cite{larsen2021geography}. This fluvial synchrograms depicted the decay in synchrony over Euclidean, watercourse, and flow-connected distances \cite{larsen2021geography}. Synchrony was higher in populations connected by direct water flow and decreased faster with the Euclidean and watercourse distances, highlighting the extent of spatial patterns of synchrony that emerge from dispersal \cite{larsen2021geography}. Other approaches like the ones of \cite{larsen2021geography, terui2018efficacy, kelly2015update, gilarranz2015inferring, holland2008strong} are examples of investigations focused on the effect of the network topology (SC) on the synchronous dynamics of nodes (FC) (SC/ FC relationships).
Representations of synchronous functional connectivity go beyond the movement-based approaches and can include models using the input-output analyses described in the ENA section (i.e., species interactions models quantifying predator-prey, mutualistic, or competitive relations \cite{bastolla2009architecture}; species-resource interactions models quantifying consumer-resources relations \cite{hajian2021viewing, lotka1925elements, volterra1926variazioni}, food webs models that trace energy movement \cite{brechtel2018master, dame1981analysis,rudolf2011stage} and nutrient cycling models \cite{christian2003network}. In competitive consumer-resource systems, consumers can overlap their diet and resources interact with one another, which makes it possible to be visualised them as coupled oscillators \cite{hajian2021viewing, Winfree1967}. To explain the dynamics of communities in these systems, Hajian-Forooshani and Vandermeer \cite{hajian2021viewing} applied the enduring Lotka and Volterra equations \cite{lotka1925elements, volterra1926variazioni} and the Kuramoto model \cite{kuramoto1975self} in a simplified three-oscillator system. In this system composed by three consumers and three resources, they measured two distinct types of coupling: trophic-coupling (the strength of cross-feeding) and resources-coupling (strength of competition between resources) \cite{hajian2021viewing}. Given a persistent oscillator in the Lotka-Volterra formulations, trophic-coupling implied eventual synchrony (all oscillators are in the same point in circle space) and resource-coupling implied asynchrony \cite{hajian2021viewing}. The simulations in both of the two models, had similar results, suggesting that coupled oscillators and the application of the notions of the Kuramoto model can provide theoretical contributions on ecosystem and community’s organization \cite{hajian2021viewing}.

To illustrate the idea of sequential activation in freshwater systems, we consider examples using random walks. Random walk is the most common approach to simulate animal’s movement and can be considered as sequential FC, since the sequential steps of their dispersal can provide valuable information about the network structure. For theoretical studies that model the distribution of local species persistence in time, random walk without drift is the simplest baseline demographic model \cite{Rinaldo2018}. In originations and extinctions models working with macroevolutionary timescales, the abundance of a species in a node has the same probability of increasing or decreasing by one individual in each time step \cite{Rinaldo2018}. Then, the increase of one individual will represent the colonisation of a free site by an individual of a new species in the system, or a randomly sampled individual within the community \cite{Rinaldo2018}. An assumption of this model to account for limited dispersal effects is that only offspring of the nearest neighbours of the dying individual are allowed to colonise the empty space \cite{Rinaldo2018}. Additionally, the local extinction corresponds to a first passage of a random walker equal to zero, leaving a persistence time distribution following a power-law decay with exponent 3/2 \cite{Rinaldo2018}. A simpler alternative model will be the stochastic patch occupancy model (SPOM) that describes the presence/absence of a focal species in a node (simulates only colonisation and not colonisation-extinction dynamics) \cite{chaput2017network, hanski2003metapopulation}. Here, in a given structural river network with discrete habitat patches as nodes, each node has a probability to be colonised by species belonging to the regional species pool \cite{chaput2017network}. At the starting point of each simulation a sequential colonisation process starts. From initially occupied nodes, or initially introduction sites of the new species, the empty patches can become occupied in a sequential manner (successive snapshots). The potential occupancy of a node will be dependent on a chain of colonisation events and the presence of unoccupied nodes within a certain range (only empty nodes could become colonised). 
Although the SC/FC relationships implied by the aforementioned studies are different from the ones described in this paper. The dispersal of animals (which serves as sequential FC) finally determines the structure of the network and in the approaches of \cite{chaput2017network} and \cite{Rinaldo2018} this one is built using time-ordered graphs (i.e., continuous-time Markov chain for \cite{chaput2017network}. The main difference is that this provides a time resolved view of the dynamics. However, in the approach of the current paper, the time is eliminated, by suggesting the time-average view on dynamics. Bridging over these two types of time views on dynamics requires further investigation on how the FC that derived from the temporal graphs contributes to the time average information of dynamics.
Ecological applications of dynamical approaches, like the ones mentioned above, and the classification of the two types of FC addressed in this paper bring new perspectives to assess functional connectivity in freshwater ecology. Additionally, evaluating the SC/FC relationships can highlight the importance of specific nodes in facilitating the overall colonisation processes, which can help to estimate a number of effective reserves necessary to achieve a particular conservation goal \cite{arumugam2021tracking, Gilarranz2015, Larsen2021, Luque2012, Moore2015, Terui2018}.

\subsection{Application to Systems Biology}
In Systems Biology, we find many instances, where a distinction between simultaneous and sequential events in networks is made, even though the terminology of 'functional connectivity' is rarely used. On the level of \textit{gene regulation}, for example, temporal programs structurally implemented via single input modules \cite{shen2002network}, leading to a 'just-in-time' production of proteins for specific biological functions in a bacterial cell \cite{zaslaver2004just} is an example of a contribution to functional connectivity based on sequential activation. Note that here the unweighted graph would lead to a misleading relationship between structural and functional connectivity, as structure in the \textit{unweighted} graph would suggest a co-activation, rather than a sequential activation. The latter, in fact, is implemented via distinct weights from the regulator to the target genes or operons. 
As we see in all the case studies presented here, such details matter, when bringing these abstract concepts to a specific domain of application. 

Another example is the sequence of events during the yeast cell cycle, which is hard-wired into the corresponding gene regulatory network \cite{chen2000kinetic} and can be understood using simple, discrete cellular automata-type models \cite{li2004yeast,davidich2008boolean}, namely Boolean network models \cite{bornholdt2005less}.

A more common approach in Systems Biology addressing the relationship of gene expression (or transcriptome) data -- simultaneous measurements of gene activity via high-throughput technologies -- and the underlying regulatory network is network inference. This approach summarizes a range of statistical methods to infer the regulatory network from expression profiles \cite{marbach2012wisdom,le2015quantitative}. It should be noted that this approach already starts with assumptions about SC/FC relationships, in particular, that indeed structural connectivity can be reconstructed from observations of the dynamical states \cite{HuettLesne2021}.

In the case of \textit{metabolic networks} -- bipartite graphs metabolites and biochemical reactions available in a cell, where reactions can be represented either by enzymes or by the genes encoding these enzymes \cite{jeong2000large,Beber:2012ix,palsson2015systems} -- the activity levels of genes encoding enzymes can be thought of as a representation of the metabolic state of a cell. 
These activity levels are given by transcriptome data. Statistical analysis of transcriptome data (either by repeated measurements -- replicates -- or by contrasting the cellular condition with a baseline or reference conditions (e.g., mutant gene expression levels with wildtype gene expression levels) leads to sets of genes characterizing a cellular state, for example, the sets of 'upregulated' and 'downregulated' genes or the sets of genes with 'high' or 'low' expression levels within a large set of samples. A possible definition of functional connectivity then is the induced subgraph of a gene-centric projection of the metabolic network spanned by such a gene set derived from transcriptome data \cite{hutt2014understanding}. 

The connectivity of such a subgraph compared to randomly drawn gene sets is a convenient and frequently employed measure for SC/FC correlations in this context \cite{sonnenschein2012network,knecht2016a,perrin2020lipogenesis,nyczka2021inferring}, as it addresses the statistical question, how clustered such a gene set (representing functional connectivity) is within a given (metabolic) network (representing structural connectivity). 

The conceptual model of functional connectivity behind such an investigation is that of synchronous activity. 
Distinguishing between the two types of functional connectivity is challenging, given the current state of 'omics' data in Systems Biology, due to the lack of suitable time-resolved data. 

This observation is further underlined by the fact that also predictive theories of genome-scale metabolic activity (e.g., flux-balance analysis, \cite{orth2010flux,o2015using}) are based on a steady state assumption. In order to discriminate between co-activity and sequential activity, one needs to resort to time-resolved models, typically based on ordinary differential equations (ODEs), which however due their often huge number of required parameters are restricted to single pathways or other suitably defined other cellular subsystems, rather than the scale of a whole cell. 

Beyond these two basic situations characterized by a gene regulatory network and a metabolic network as structural connectivity, respectively, there are many other examples of sequential and synchronous usages of a given 'hardware' in Systems Biology. Metabolic control analysis \cite{fell1992metabolic,cascante2002metabolic} relates the distribution of control in biochemical networks to their structure. Protein interaction networks \cite{oughtred2019biogrid} summarise, how selective binding patterns (structural) and protein complexes (the dynamic assemblies to execute biological function) are interlinked. Fermentation processes (e.g., in cocoa fermentation) often rely on a sequential activation of microbial populations \cite{moreno2018mathematical}.

Summarizing the SC/FC situation in Systems Biology in a qualitative form, one can conclude that gene regulatory networks lean towards sequential activation, while protein interaction networks functionally lean towards co-activation (protein complexes) and metabolic networks may display aspects of both (steady-state activity vs. metabolic pathways).

\subsection{Application to Neuroscience }

Neuroscience is one among the first disciplinary fields in which the need to formalize notions of SC and FC was felt. Perhaps, this was due to the fact that network descriptions in neuroscience go beyond a mathematical representation but correspond to an actual, concrete reality: neural circuits are networked systems, with their “reticular” (from Latin for “little net”) nature debated since at least the turn of last century \cite{jones1999golgi}. Network nodes can be, depending on the scale of observation, individual neuronal cells (at the micro-scale), populations involving thousands of neurons (at the meso-scale), up to entire brain regions (at the macro-scale). The relations defining links are different depending on the considered type of connectivity and are defined both in terms of anatomy of wiring and of information exchange.

It is natural to consider SC in neuroscience as the description of anatomical connections physically existing between network nodes: individual synaptic connections forming electrochemical junctions between the outward axons and the inward dendrites of different neurons (within volumes $<$ 1 mm$^3$, already containing $\sim$10$^4$-10$^5$ neurons); or bundles of long-range connection axons coupling together smaller or larger groups of neurons, separated by varying distances ($\sim$1-10 mm for mesoscale circuits up to $\sim$10-100 mm for macroscale, brain-wide networks). At all these scales, one usually refers to the compilation of all structural connections between probed network nodes as to a connectome \cite{sporns2005human, seung2011towards}. Different techniques must be used to extract SC information at different scales, even if a systematic review of them is not possible here. It will be enough to mention that we dispose today of whole matrices of SC for rodent, nonhuman and human primate brains \cite{kotter2004online, hagmann2008mapping, chiang2011three, van2012human, oh2014mesoscale, markov2014weighted, van2016comparative}, as well as detailed microcircuit reconstructions \cite{helmstaedter2013connectomic, markram2015reconstruction, eichler2017complete}.

Studies of SC in neuroscience often revolve around: the search for general architectural \cite{ercsey2013predictive} or wiring optimisation \cite{cherniak1994component, chklovskii2002wiring} principles in connectivity; or the identification of characteristic motifs of connectivity that are over-represented with respect to chance-level \cite{sporns2004motifs, song2005highly} and special structures such as dense clusters at the micro-scale level \cite{perin2011synaptic} or cores and “rich-clubs” at the macro-scale level \cite{hagmann2008mapping, van2011rich}. Recently, attempts have also been made to use topological data analyses techniques \cite{sizemore2018cliques, bassett2017network} to characterize the “shape” of networks without beyond the limitations of graph theoretical descriptions, which are exceedingly emphasizing strictly local or strictly global aspects but are deficient in capturing intermediate structures at arbitrary meso-scales. Other lines of research aim at linking specific structural motifs to specific functions: as in the case of specific arrangements of positively weighed excitatory connections and negatively weighed inhibitory connections allowing modulations of perception \cite{carandini2012normalization} or of specific patterns of interconnection between cortical layers at different depths in the tissue thickness allowing the regulation of sensory and predictive information flows \cite{bastos2012canonical}. Finally, many efforts have been devoted to identify SC alterations that may be indicative of developing and progressing neurological or psychiatric pathologies, and thus serve as diagnostic or predictive biomarkers \cite{van2011rich, fornito2015connectomics}. However, a comprehensive survey of all these applications largely transcends the scope of this review.

Not only are neural network nodes (neurons or populations) wired together by living cables, but messages are continually exchanged along these cables. All information processing related to our perception, our cognition and our behaviour is generally believed to arise from the exchange of “spikes” -- propagating pulses of electric depolarization of the cell membrane, able to elicit neurotransmitter release at synaptic terminals -- between synaptically connected neurons. Such spikes, individually or grouped in more complex spatiotemporal patterns, represent “codewords” encoding information about external and internal worlds in still largely unknown languages. Streams of spike-encoded information are thus copied, transferred and merged between system’s components linked by SC, assembling into emergent neural algorithms which ultimately underlie functions and behaviours \cite{marr1976understanding}. These computations are highly distributed and the communication between system’s units that they involve can also be seen as giving rise to networks, but this time of functional rather than structural nature. Two units are thus defined as functionally connected if they “interact”. The problem becomes thus to operationally define how an “interaction” can be pragmatically measured from the observations of how coordinated neural activity unrolls through time.

Some measures of FC define “interaction” as “synchrony” between activity fluctuations and modulations. This is the case for instance of so-called resting state FC \cite{fox2007spontaneous}, describing linear covariance between the fluctuations of different brain regions during unconstrained mind-wandering, as revealed by functional MRI (fMRI). Such metrics of connectivity are akin to the first form of FC previously described. Given the remarkable oscillatory components present in neural activity and simultaneously at different frequency bands \cite{buzsaki2006rhythms, wang2010neurophysiological} analyses of synchronisation-type FC in neuroscience are often conducted in the spectral domain, tracking coherence and phase-locking \cite{varela2001brainweb}. Individual neurons are not necessarily oscillating and can keep firing in irregular manner, nevertheless neuronal populations can collectively oscillate, because of the interplay between excitatory and inhibitory currents within local recurrent microcircuits \cite{brunel2003determines}. Such collective oscillations produce periodic modulations of the excitability of neurons within large populations, so that efficient transmission between synaptically coupled populations can occur only if their respective local oscillations are suitably aligned in phase (“communication-through-coherence” hypothesis \cite{fries2015rhythms}). Thus, two neuronal populations that are structurally connected may become functionally disconnected if their oscillations are e.g. in antiphase and spikes emitted in proximity of the senders population’s oscillation peaks reaches postsynaptic neurons at the throughs of the target population’s oscillations and hit thus against a wall of locally-generated inhibitory blockade, preventing information carried by input spikes to be transduced into output activity. Under this hypothesis, it is thus possible to flexibly “switch on and off” FC on top of a static SC link, just by adjusting the relative phase of the sender’s and target’s oscillations. A natural generalization of linear correlation from the time to the spectral domain when dealing with oscillatory neural activity is inter-regional coherence or phase synchronisation \cite{varela2001brainweb}. Coherence in the gamma band (40-100 Hz) between frontal/prefrontal and sensory regions is known for instance to be boosted in sensorimotor coordination or attention \cite{roelfsema1997visuomotor, fries2001modulation}. Furthermore, FC can also be established between populations oscillating at different frequencies via nonlinear cross-frequency coupling \cite{canolty2010functional}. Not only cognition, but also pathology can perturb coordinated neural oscillations and the associated FC \cite{uhlhaas2012neuronal}, but once again a detailed coverage of the use of oscillation analyses for biomarking goes beyond the limits of the present work.

Other measures of FC go beyond mere synchrony or correlation -- beyond the first form of FC -- and attempt reflecting actual causal influence. Unlike correlation which is symmetric and thus give rise to undirected graphs of FC, measures of causal interdependence between time-series of neural activity give rise to directed networks. The name of “effective connectivity” has been sometimes used in neuroscience \cite{friston2011functional} to refer to FC measures that track causality, but here we keep naming connectivity of functional type all connectivity relations that do not express anatomical interconnection. A very simple way to account for the direction of interaction can be to assess the temporal precedence of the “causing” on the “caused” fluctuation. This can be achieved for instance by using lagged cross-correlation or mutual information rather than zero-lag correlation (see e.g. \cite{liu2006spatiotemporal}), since the effect cannot precede the cause.
In this sense, these metrics are related to the second form of FC, reflecting sequential activation, as previously discussed. However, the correspondence is only partial, in this case and unlike for neural FCs of the first form. If exact sequences of neural activation can be produced by neural architectures known as “synfire chains” \cite{abeles1991corticonics}, they have been only rarely sought for in actual recording and neuroimaging data \cite{ikegaya2004synfire, mitra2015lag} and not been put in relation with notions of functional coupling. Directed FC measures in neuroscience are defined operationally in terms of time-series based statistical metrics, rather than in terms of explicitly dynamic considerations. Thus, while in many cases the statistically -- inferred directed FC goes, e.g., from the phase-leading to the phase-lagging neuronal population \cite{battaglia2012dynamic, bastos2015communication}, i.e. respects a sequentiality criterion, in other cases the relation can be inverted, reflecting non-linear interactions between populations, as anticipatory synchronisation \cite{carlos2020anticipated} or heterogeneities in internal synchrony levels \cite{deschle2019directed}. 
More explicitly, causality could be captured: by the detection of remote effects on distant regions triggered by interventions in local regions (as in “Dynamic Causal Modelling” \cite{friston2011functional}); or, by showing that consideration of the past activity of a putative causal source region improves the prediction of the future activity of a target region, as in Granger Causality analyses of neural time-series \cite{brovelli2004beta, ding200617, gregoriou2009high,brovelli2015characterization}. Importantly, Granger causality can also be spectrally decomposed \cite{dhamala2008estimating}, allowing to detect the contribution of different oscillatory components of neural activity to inter-regional causal influences. It has thus become possible to observe that causal influences in different directions can be mediated by oscillations in different frequency bands, e.g. in the gamma-band ($\sim$40 Hz) for bottom-up and in the beta-band ($\sim$20 Hz) for top-down information exchange between prefrontal and visual regions \cite{bastos2015visual,michalareas2016alpha}.

More recently, emphasis has been put on the fact that FC networks are not static but change in very flexible way through time, i.e. they are better described as temporal networks \cite{holme2012temporal}. The new term of “chronnectome” has been introduced to stress how, beyond analyses of the static functional connectome, explicit consideration of the spontaneous reconfiguration dynamics of FC along time may help better discriminating cohorts of subjects and patients, by disentangling temporal from inter-subject variability \cite{calhoun2014chronnectome}. At the macro-scale, different FC networks are sequentially recruited along the unfolding of cognitive tasks, potentially signaling different neurocomputational steps \cite{ioannides2007dynamic, brovelli2017dynamic}. Spontaneous resting state FC networks wax and wane in a seemingly stochastic flow which is not randomly structured but display characteristic long-term memory \cite{battaglia2020dynamic} and whose rate of reconfiguration and degree of temporal structuring predicts cognitive performance at the single subject level \cite{braun2015dynamic, lombardo2020modular}. At the micro-scale as well, the emergence and dissolution of transient synchronous assemblies of cells within hippocampus and enthorinal cortex can be modelled with temporal network descriptions \cite{pedreschi2020dynamic}. Future studies be needed to understand whether this complex FC network dynamics can be seen as a measurable fingerprint of ongoing neural computations linked to functional behaviour \cite{clawson2019computing}.

The definitions of SC and FC given in the previous subsections are in principle completely independent: one could indeed assess the existence of FC based on the analyses of multivariate neuronal activity time-series without knowing anything about the underlying anatomy and SC. This scenario of a “perfect separation” between SC and FC is obviously unlikely. An equally naïve scenario dominates however the discussion of many articles in the literature in which a structural cause is necessarily sought for to explain any change arising at the level of FC. In reality, the flexibility of FC on behavioural time-scales very fast with respect to physiological processes reshaping SC (at least at the meso- and macro-scale) already suggests that FC cannot just be a passive mirror of the underlying SC. We have previously proposed that FC is the measurable by-product of underlying collective dynamics \cite{battaglia2012dynamic, kirst2016dynamic}. In this proposed theoretical view \cite{battaglia2014function,battaglia2020functional}, alternative modes of system’s dynamics, or states within the “dynome” \cite{kopell2014beyond} -- or dynamical repertoire \cite{ghosh2008noise} -- of a system would give rise to alternative FC configurations on top of a same underlying SC (functional multiplicity).

Analogously, circuits with very different SC but that give rise nevertheless to equivalent dynamical modes – a property known in systems neuroscience as functional homeostasis \cite{marder2006variability} – would give rise to similar FC (structural degeneracy) \cite{marrelec2016functional}. An example of degeneracy can be found in simulations of neuronal cultures in vitro, in which high clustering of FC is invariantly found because of collective network activity bursting, independently from the simulated culture’s SC being weakly or strongly clustered \cite{stetter2012model}. 

It is important to stress that, in our view, FC manifests collective dynamical modes of the entire neural system considered more than properties of node-specific dynamics. This is made evident in studies that attempts to predict large-scale FC in spontaneous resting-state activity conditions starting from simulations of SC-connectome-based simulations. In these cases, a good fit between simulated and empirical FC is obtained only when the model is tuned to operate close to a critical point of dynamic operation \cite{honey2007network, deco2011emerging}, indicating that FC manifests a peculiar dynamical regime, certainly shaped but not fully constrained by structure. For instance, waves \cite{muller2008organization, moretti2020link} or “connectome harmonics” \cite{atasoy2016human} shape large-scale coordinated activity and, hence, FC. The importance of being tuned into specific dynamical regimes to account for the qualitative features of large-scale coordinated activity have also been confirmed by minimal models, with reduced realism but enhanced possibility to rigorously understand mechanisms \cite{garcia2012building, deco2012anatomy, haimovici2013brain}.
A predicted consequence of this hypothesis is that local perturbation to individual nodes within a neural network may induce a network wide reconfiguration of FC, including of remote nodes not directly connected to the perturbed node \cite{kirst2016dynamic}. Furthermore, the effects of perturbation could be dependent on FC, rather than on SC, as, in a nonlinear system, a same perturbation will yield different effects in different dynamical states and it is FC to be state-dependent. Once again, computational simulations of virtual brain models informed by empirical SC information but augmented with nonlinear brain dynamics, confirm the validity of our prediction \cite{papadopoulos2020relations}. Virtual brain models tuned to regimes maximizing the degeneracy of their “dynome”, sampled via a noise-driven exploration, also succeed in reproducing qualitatively the switching “chronnectome” observed in resting state fMRI \cite{hansen2015functional}. Computational modelling provide thus strong evidence in favour of our hypothesis of flexible FC being the by-product of a complex dynamical system, whose behaviour is constrained but not fully determined by the underlying SC. In other words, function follows dynamics, not structure.

\subsection{Application to Social-Ecological systems}

SES are complex adaptive and multilevel (polycentric) systems attributed with interplays between human and non-human entities (nodes) at spatial and temporal scales \cite{folke2016social}, through the metabolic flows of material and energy (links). The concept of “social metabolism”, taken from cellular metabolism, is central in the study of SES \cite{martinez1987ecological}. network analysis has increasingly been used to study coupled, or social ecological systems \cite{bodin2011social, barnes2017social, bodin2012disentangling, sayles2019social, prell2017uncovering}. Here, the SES is often depicted as a multilevel social-ecological network (SEN), where social/human actors comprise one network, natural entities a separate network, and flows are captured between and within each network level. Such multilevel networks are modelled via a stochastic environment, such as a Multilevel ERGM \cite{wang2013exponential}. Here, micro-configurations are specified, consisting of actors and/or entities from either or both the two networks, such as the tendency for two social nodes to share a coordination tie, when both nodes are likewise linked to the same natural resource. These micro configurations are then modelled, alongside other competing tendencies (such as the general tendency of a network to exhibit transitive closure), to test hypotheses linking SEN patters to sustainable (or unsustainable) management practices. 

\subsubsection{Social networks}
The standard SC/FC approaches are usually uncommon in social networks because the distinction between the 'hardware' (structural connectivity) and the dynamics (functional connectivity) is less clear than in other fields. However, measuring the (often rapid) flows along the edges of a more stable (slowly changing) network, as the example of the competence perceptions and the everyday information exchanges in a company shows, could be an interesting perspective for future research approaches, incorporating two theoretical frameworks: social systems theory \cite{saenz2007democracias, luhmann1984soziale} and social network analysis \cite{prell2012social,scott2013social, jansen1999einfuhrung}, e.g. in the context of relational events models \cite{stadtfeld2017interactions}.

Both approaches are debating how the internal function (of networks or systems) can produce emergent properties that transform the structure and vice versa. However, in social science, these debates are not without tensions (i.e., between structure and agency) and criticisms (i.e., by more conflict-oriented approaches).
We argue that the overlap between 'system' and 'network' could be helpful for SC/FC in social science, specifically to understand the connections between actors embedded in different social subsystems and how underlying network topologies among those actors impact the subsystems (for example, economy and democracy), and vice versa. Signed contracts between public institutions (PI) and private companies (PC) can serve as an illustrative example. The outcome of a relational analysis of public procurement is a multilevel, multirelational, two-mode network of business-government connections, whose nodes and relations are embedded in different social subsystems: the economic system, the political system and the State. Here, the main focus is the SC/FC relationships that emerge from the analysis of the procurement network.

FC$_{\mbox{{\scriptsize seq}}}$ is the longitudinal and dynamic procurement process, analysing the sequential configurations to understand why a PI is issuing a contract, whether to one and not to another PC. Some of these companies could be important market leaders or potential corrupters. FC$_{\mbox{{\scriptsize seq}}}$ is especially significant in the case of private actors, as we assume that very outstanding degree-peaks of a few companies could be evidence of a corrupted network dynamic. Companies with an extraordinarily high number of contracts in a short period may have extraordinary political influences (i.e. interlocks or bribes). Both the relational positions and the dynamical peaks could correlate directly to structural network transitions (collapse or even fragmentation) and also to system-related implications, such as the resource distribution in the economic system or the decision-making process in the State.
FC$_{\mbox{{\scriptsize sim}}}$ is the specific linkage-configuration at each time step: nodes that have active links to the same node(s) are co-active. The co-activation through shared links is changing in each time step when a new link is created, and an existing link is decaying. FC$_{\mbox{{\scriptsize sim}}}$ applies to PI and also to PC, and can be seen as an indicator for strong relational positions of other nodes (in the opposing type) in the network. For example, many co-active institutions are a ‘pointer’ to influential companies, whether important market leaders or potential corrupters.

\subsubsection{Ecological networks}
Ecological Network Analysis is a systems-oriented methodology developed by ecologists to understand whole-system dynamics and properties \cite{fath2007ecological, ulanowicz1983identifying}. This methodology is based on network and information theory and derives itself from input-output analysis, modelling ecosystems as a set of nodes and ties (vertices, edges) \cite{leontief1936quantitative, miller2009input}. Under this framework, species, aggregation of species into functional groups, or non-living resource pools are taken as nodes while the exchange of material or energy between species is taken as edges. In addition, ENA methodology has also been widely applied to analyse direct and indirect exchange of energy and carbon emissions between economic sectors at urban/country level from a system perspective \cite{chen2015urban}. This methodology is useful to evaluate system properties such as cycling index, total throughflow and relational interactions by pair-wise components in the system through thermodynamically conserved transactions of a chosen currency \cite{fath1999quantifying}.

Although not explicitly using the language of structural connectivity and functional connectivity, ENA internal logics resemble the one in connectivity science \cite{turnbull2018connectivity}. Under the ENA framework, SC is defined by the number and position of functional groups - species, aggregation of species or economic sectors - forming the nodes and their flows of material and energy between them (edges) \cite{fath2007ecological}. This set of arrangements define the network architecture or network topography and, therefore, the “hardware” on which dynamic processes take place, normally represented with an adjacency matrix \cite{dame1981analysis, fath2007ecological}. Ecosystems are open, thermodynamic, far-from equilibrium systems, which implies that they require continual input flow of high-quality, low-entropy energy \cite{fath2004distributed}. Once energy enters the system, it is the structural connectivity that defines the system’s overall dynamic flow-storage patterns \cite{fath1999quantifying}. ENA is applied to steady-state systems, therefore capturing, in a snapshot, both the structural connectivity and the cumulative behaviour of a given highly dynamic network. 

Now, we turn to the functional connectivity under the ENA framework, employing a basic input-state-output model frame. As open systems require continual input, an ecosystem is sustained by the dynamic co-activation pulses entering across the boundary. In nature, these pulses could be seen as the solar energy received by the primary producers (multiple individuals or multiple species depending on scale). In this manner, we interpret co-activation as nodes sharing a functional similarity, such as trophic level, and thus being charged simultaneously. This is different than viewing co-activation as two or more attributes to align for activation to occur. The latter may not have a direct analogy in ENA. In this case, the input of energy is simultaneous to several nodes due to their inner characteristics (e.g., the all belong to the same trophic level). Once co-activation occurs, the energy/material flows sequentially from node to node. Although ecological networks have complex connection patterns including cycling, each individual sequential pathway can be “decomposed” and identified as a unique carrier of energy matter from initial activation to final dissipation beyond the system border.
These energy flows are the base of all exchanges and form the model structure encompassing a diversity of nodes and trophic levels. The sequential activation is captured along these cascading indirect pathways from the initial co-activation pulse. Therefore, the most straightforward way to visualize functional connectivity based on the sequential activation of nodes is with a linear food chain. The initial input of energy triggers the sequential activation of nodes down the food chain, whereas each component is dependent on the previous for its flow source\cite{fath2004distributed}. Eventually, as the initial pulse travels throughout the many networked pathways, it is dissipated, its useful energy spent, coming to rest outside the system boundary (as higher entropy) and completing the input-state-output triumvirate. Ecological network analysis can expose some of the interesting properties that emerge in the state based on those input-output relations.

What is particular about ecological, and therefore also SES, systems and networks is that one major element conditions both their architecture and their functional connectivity over the long run: net energy (as an indicator of low entropy) \cite{hall2009minimum, georgescu1975energy, tainter1988collapse, odum2007environment}. As long as there are high levels of net-energy, connectivity (and therefore complexity in terms of nodes and functions) can increase, as new “agents” or “elements” are attracted to the system or drawn into it by existing agents.

Nodes that happen to (or managed to) control large amounts of net-energy flows, can leverage their relative position in the network and exert power over other network members. This means, such agents then have substantial power to adapt the network architecture to their own preferences e.g. to increase their relative power \cite{hornborg1998towards}. It acts collectively on major sets of nodes, thus it contributes to synchronous F$_C$ and it can also trigger cascading effects within the network contributes to sequential F$_C$. Power enables agents to exert a certain level of control over other agents and even allows them to eliminate or add other agents or nodes. In particular they may control the distribution of flows as they move through the system.

\subsubsection{Social-Ecological networks}

A rich body of literature on social-ecological system analysis focuses on the structures and patterns of interdependent social and ecological interactions (SC), which are further associated with phenomena of interests like cooperation and conflict \cite{barnes2019social,janssen2006resilience, sayles2019social,sayles2017social, schluter2019capturing}. More specifically, it investigates the actor-to-actor relationship in the social system, the ecological component-to-component interdependencies in the ecological system, as well as the actor-to-component relationship across the social and ecological system \cite{bodin2020reconciling}. Altogether it forms a multilevel network configuration made of nodes and links between different system entities. In terms of SC/FC relationships, one line of research is exploring how certain social-ecological system configurations can facilitate successful adaptation and transformation in SES to address resource management challenges\cite{barnes2019social, nelson2007adaptation, newman2005network}. Both adaptivity and transformability are critical elements of resilience study, describing the capacity of the interdependent social-ecological systems dealing with unknown or unforeseen shocks \cite{folke2010resilience, walker2004resilience}.

Although using different terms, other lines of research have identified two types of cascading effects that connect various regime shifts, the directional and bidirectional links\cite{da2018product, kimmich2013linking}. One is called the domino effect that reveals a one-way directional dependence \cite{hughes2013attenuation}. We argue that it fits more with sequential functional connectivity due to the fact that the feedback from one regime shift affects the drivers and outcomes of another regime shift. While the other one is termed hidden feedback, showing a self-amplifying/damping bidirectional cycle \cite{liu2015global, schluter2019capturing}, which we argue is more of a synchronous connectivity nature. 

Various analytical frameworks have been applied to capture the process of co-evolution, such as the MuSIASEM (Multi-Scale-Integrated Assessment of Societal and Ecosystems Metabolism) framework. From a MuSIASEM perspective flows of material and energy move through a system (or network) in order to fulfil certain societal functions. We argue that it departs from a set of known structural connections (e.g. the mix of primary energy sources for a society and its end uses) and then tries to describe functional connectivity of a central element of a network structure by using ratios that are composed of both, a flow and a fund element. Flow is the element that either disappears over the duration such as primary energy or appears by the end of the duration like the product, while fund can be seen as a converter that transforms input flows into output flows during the enter-exit duration e.g. labour, land or machinery. Moreover, funds are impermanent structures whose existence depends on the availability of flows \cite{georgescu1975energy}. These ratios give (among other things) information about the relative power of nodes/agents in multi-level networks. High rates of metabolized energy provide increased power to (1) control and both create and synchronously co-activate many nodes/ agents in a network (“hierarchy-dependency effect”) and (2) to influence sequential activation by controlling flows (“controlling-the-tap-effect”). MUSIADEM tries to provide measures and indicators for such relations, in order to guide the transformation towards a Post-Carbon society. 

Another concrete example of a SES here is the global commodity trade system connecting resource extraction and final demands. Here nodes are the trading partners such as cities/countries/regions (at various jurisdiction levels), which can be linked through flows of products, material, monetary value, and environmental footprints. Altogether, the established static trading structure with complex interactions constitutes the network architecture (SC). For instance, in the palm oil trading market, Indonesia and Malaysia have been the main producers, exporting products to countries like the EU, China and India. The identified relational structure between the countries is the SC. On the other hand, network dynamics (FC) describes the dynamics of the flow (i.e. the quantity of trade; the environmental footprint) embedded in the relational patterns. Input-Output Analysis \cite{leontief1974environmental, miller2009input} has been widely used to capture the input flows among each sector of trade partners in the network. The flow dynamic in the IO table is rather synchronous, in the sense that it is the market interaction where price co-activates both supply and demand sides. For instance, with the EU passing a stricter sustainability regulation while importing palm oil, big producers like Indonesia tend to export more of their products to less regulated markets like China. The network structure remains the same, yet the flow dynamic changes synchronously as driven by the market price (i.e. higher standards will increase the production costs, thus the price will rise accordingly). Sequential activation cannot be modelled using IOA, as it is more like a snapshot of an economic system in a given moment in time. In fact, this static nature of IOA is often criticised as one of its major shortcomings that have only partially been overcome by the development of dynamic models.

Although connectivity terminology is not explicitly used here, the phenomenon of network evolution through actions of its agents (nodes) is found quite evident. The theoretical framework developed in the paper regarding the distinction between synchronous and sequential events has a great potential to provide a different network perspective to understand the underlying mechanisms in social-ecological networks.

\color{black}
\section{Conclusions}
Here we have attempted to unify the broad range of SC/FC approaches within a common framework. We have reproduced key findings from the literature and extended them towards additional variations of network topology and dynamical characteristics, in order to see common properties and underlying principles and offer a deep mechanistic understanding of the major contributors to SC/FC correlation. 

Minimal models (small toy model representations of certain classes of dynamics) are helpful to explore these generic features. Our challenge here was to describe how the strengths of the two types of SC/FC correlations -- based on co-activation and sequential activation -- depend on the class of dynamics, the network architecture, the coupling and the internal dynamical parameters. We used numerical simulations to derive some universal behaviours of SC/FC correlations under changes of these system properties and to apply this knowledge to real-life systems or data.

The strength of SC/FC correlations can be shifted between the two classes -- functional connectivity based on co-activation and sequential activation -- in basically three ways: (1) Modification of network architecture (e.g., the gradual randomisation of a modular graph), (2) change in parameters of the dynamics (e.g., increasing or decreasing the noise or the coupling), and (3) a change in the temporal resolution in which dynamical data are observed (e.g., by temporally coarse-graining the observed time series).

The basic challenge of this type of investigation is that the strength of each type of SC/FC correlation depend not only on the class of dynamics, the network architecture, the coupling strength and the dynamical parameters, but also on the type of statistics that are applied. In some cases, the effect of the different statistics is so strong that changes there is noticeable change on the properties that are preserved or not. 

An important question is how to assess the reliability of the results. In order to confirm that a numerically observed behaviour of SC/FC correlations (under systemic changes) is reliable, we performed the following tests: 
(1) We vary the other system parameters slightly, in order to study the robustness of the result. 
(2) We reproduce the behaviour observed in a minimal model also in a richer representation of the same class of dynamics. 

Of course, the question is more involved on the technical level than our brief introduction hints at. There are different ways of assessing functional connectivity beyond pairwise correlations. Across all disciplines, the reliability and completeness of structural data is an important issue. In the case of brain networks on the level of cortical areas (or 'connectomes'), one issue is whether or not to regard these networks as weighted or unweighted graphs \cite{markov2014weighted,buzsaki2014log}. Furthermore, most systems, for which such correlations are of interest, will have some form of multiscale organization \cite{gallos2007scaling}. Hence, any analysis on SC/FC relationships will require selecting suitable spatial and temporal scales. On some level, we can furthermore expect that the structural network (often thought of as 'static' in the context of SC/FC correlations) will also change with time, though often on a longer time scale than functional connectivity. We can furthermore envision a co-evolution of structural and functional connectivity towards jointly ensuring a reliable functioning of the system \cite{damicelli2019topological}.

Even the small discussion of the plausibility of these stylized forms of dynamics in the context of the application domains shown in Figure \ref{fig0} illustrates how real-life complex systems contain a range of dynamical usages (functional activity patterns) of a given infrastructure (structural connectivity). It is less clear, however, that even a form of dynamics, which by definition seems to favour one type of functional connectivity (sequential activation for excitable dynamics; synchronous activity for coupled oscillators) can display strong SC/FC signals for the other type of functional connectivity, if the constellations of network architecture, coupling and dynamical parameters are right. This point is illustrated with the numerical simulations discussed here. 

We believe that subsequent investigations might employ the pattern of SC/FC correlations as a means 
of identifying from a given network structure, which type of dynamics is most plausible, i.e., which type of dynamics this network was 'built for'. Our current understanding of dynamics on networks does not yet allow for such a detailed assessment.

On the technical level, various definitions of co-activation and sequential activation are plausible, e.g. different normalizations, time delays and discretizations. We did not explore these aspects in detail. A discussion of the impact of these aspects can be found for example in \cite{messe2018toward,damicelli2019topological}.

As often with numerical investigations, some seemingly small 'design decisions' affect the results. In the SER model, for example, near the deterministic limit, longer runs do not provide more information, as the systems rapidly settles into a (periodic) attractor. Then, only a large number of short runs can reveal the underlying network architecture. The same is true for phase oscillators, which provided the coupling is high enough given a certain spread of eigenfrequencies, rapidly settle into a fully synchronised state no longer informative about the architecture of the network. Here also, transients from many runs need to be collected.

With our investigations we set out to understand which network features and which class of dynamics rather enhance SC/FC$_{\mbox{{\scriptsize sim}}}$ or rather enhance SC/FC$_{\mbox{{\scriptsize seq}}}$. As a rule, we find that modularity enhances SC/FC$_{\mbox{{\scriptsize sim}}}$, while broad degree distribution or randomness enhances SC/FC$_{\mbox{{\scriptsize seq}}}$. Increase in coupling favours high SC/FC$_{\mbox{{\scriptsize sim}}}$, while high parameter diversity tends to enhance SC/FC$_{\mbox{{\scriptsize seq}}}$. 

From the view of dynamics, excitation models tend to favour FC$_{\mbox{{\scriptsize seq}}}$, a trend we observe both with the minimal (SER) model of the excitable dynamics and with the more realistic FitzHugh-Nagumo model. In contrast, regular oscillators favour FC$_{\mbox{{\scriptsize sim}}}$, as we see with the stylized (coupled phase oscillator) model and, at a higher level of realism, with the FitzHugh-Nagumo model in its oscillatory regime. In the case of the chaotic oscillators, the choice of the coupling term used here leads to a persistant dominace of high positive SC/FC$_{\mbox{{\scriptsize seq}}}$, but for a complete view about the SC/FC strengths further investigation with other types of coupling is needed.

The conceptualisation of the synchronous and sequential activity in different application scenarios is more sophisticated than in the case of minimal models, which indicates that broader definitions of the two notions are needed. However, as we show in the second half of our investigation, the systematics extracted from investigating minimal models help us better organize the diverse findings in the application domains and thus provides a fresh perspective on dynamical processes in network-like systems in these fields. Specifically, we argue that FC$_{\mbox{{\scriptsize sim}}}$ is associated with simultaneous measurements either of the dynamical activity of nodes or of links, where the concept of 'simultaneous events' introduces a time scale, at which events are considered to be 'synchronous'. Relatedly, in terms of a more general view on FC$_{\mbox{{\scriptsize seq}}}$ , we argue that it can be seen as flow of information or materials in the system, summarizing concepts, such as influence and diffusion.

\section{Methods}

\subsection{Network topologies}
A set of abstract graphs (modular, Erd\H{o}s-R\'enyi, Barab\'{a}si-Albert, Newman – Watts – Strogatz and hierarchical \cite{ravasz2002hierarchical}) and three real life networks (neural \cite{markov2014weighted}, social \cite{cross2004hidden}, metabolic \cite{palsson2015systems}) were used to perform the simulations on. The description of the network architectures of the graphs is given below: \newline
\textbf{Modular graph}: includes 60 nodes and has density 0.23. Each graph is constructed by starting from 4 cliques, in which every node is linked with all the other nodes in the same clique. Then, edges are randomly rewired with probability p=0.23 to link different cliques. \newline
\textbf{Watts – Strogatz graph}: includes 60 nodes and every node of the graph is linked with its 15 nearest neighbours in a ring topology \cite{watts98}.\newline
\textbf{Erd\H{o}s-R\'enyi graph}: includes 60 nodes and the probability of edge creation for each node is 0.23 \cite{Erdos-Renyi} \newline 
\textbf{Barab\'{a}si-Albert graph:} includes 60 nodes. Each BA graph was created by gradually adding new nodes each one with 8 edges \cite{barabasi1999emergence}. \newline
\textbf{Hierarchical graph}: includes 64 nodes, 174 edges and it has a scale-free topology with modular structure. The detailed construction process is described in \cite{ravasz2002hierarchical}.

\textbf{Neural graph}: includes 89 nodes (cortical areas) and 676 edges derived via thresholding and symmetrisation from the 29$\times $91 connectivity matrix (inter-areal connection strength measurements) described in \cite{markov2014weighted}. An edge between two nodes is accepted if the decimal logarithm of the corresponding connection strength measurement is above the threshold value $10^{-3}$ (see Supporting information for more details).

\textbf{Metabolic graph}: includes 72 nodes (metabolites) and 486 edges \cite{palsson2015systems}. We use the SBML model 'e-coli-core' from the BIGG database (bigg.ucsd.edu) and extract the stoichiometric matrix $S$. The adjacency matrix of the metabolite-centric metabolic network shown in Figure 1 is then obtained by mapping all non-zero entries in $SS^T$ to $1$, where $S^T$ is the transpose of $S$.

\textbf{Social graph}: includes 77 nodes (people) and 875 edges \cite{cross2004hidden}. It is an undirected graph and an edge between two nodes is created, if one's knowledge about the skills of others within the company exceeds a threshold equal to 5.0
(see Supporting Information for more details about the real life networks).

\subsection{Topological changes}
In the first instance, three initial graphs that have distinct structure were randomised or rewired in different proportions, such as the ratio $\dfrac{NoChanges}{NoEdges}\simeq 0.11$ corresponds to a percentage of $10\%$ of randomisation / rewiring process. Thus, for the modular and the regular graph every $10\%$ of randomisation / rewiring process corresponds to 50 swaps/rewiring changes of edges. The degree of the nodes is preserved and only the structure of the network changes. For the hierarchical graph, 20 swaps of the edges for every $10\%$ of randomisation are enough to end up with a scale-free graph, whose modularity is completely destroyed. As a consequence, the randomised network retains its degree distribution and the presence of hubs, but without the embedded modularity that initially had, similar to a scale-free topology as the preferential attachment model from \cite{Barabasi:1999uu}.
The modular and the hierarchical networks were randomised, according to the Markov chain algorithm \cite{maslov2002specificity}: pairs of randomly selected edges are swapped, providing no self-loop or multiple edges between two nodes are created. 
The rewiring process was performed on the Watts – Strogatz model according to the scheme from \cite{watts98}: a randomly selected link was destroyed and a new one was created between one of the two nodes and a randomly selected one; the requirement of self-loops and multiple edges between two nodes must be, also, satisfied. During the rewiring process and before we end up with an Erd\H{o}s-R\'enyi graph, the network passes through a 'small world' regime \cite{watts98}.

\subsection{SER model}
 The models we used to highlight the two classes of functional connectivity cover a range of different types of dynamical processes: excitable dynamics, regular oscillations and chaotic oscillations. 
The SER model, a simple cellular automaton model of excitable dynamics, acts on discrete time and the update rules are simultaneously applied as follows to go from the state at time $t$ to the state at time $t+1$: (1) A node in the susceptible state ($S$) changes into a node in the state of the excited nodes ($E$), if one or more of its neighbours are excited. Alternatively, a node can go from $S$ to $E$ in a stochastic way with a given (usually small) rate of spontaneous excitation, $f$. (2) A node in the excited state ($E$) changes into a node in the refractory state ($R$). (3) A node in the refractory state ($R$) changes into a node in the susceptible state ($S$) in a stochastic way with a given refractory probability $p$. This model has been originally studied as a model of self-organized criticality \cite{drossel1992self} and later been applied to address abstract questions of excitable dynamics on graphs \cite{muller2006topology,Garcia2014,fretter2017topological}, as well as topics in neuroscience \cite{muller2008organization,damicelli2019topological}. 
In the deterministic limit, $p=1, f=0$ the contribution of the 3-cycles affects significantly the collective dynamics \cite{Garcia2012,messe2018toward}. 
Due to its discreteness in time and states, in the SER model co-activity and sequential activity of the nodes can be defined in a parameter-free way: Each node can be found in one of the three states $x_{i}(t) \in \{S, E, R\}$, however in the analysis of SC/FC relationships we only distinguish two states:
\begin{equation*}
c_{i}(t) = \begin{cases}
1 &\text{$x_{i}(t)=E$}\\
0 &\text{$x_{i}(t) = S \hspace{0.1cm} or \hspace{0.1cm}  R$}
\end{cases} .
\end{equation*}
Separating the nodes into the two categories (active or inactive) is a convenient way to define the two classes of functional connectivity. The co-activation matrix is
 \begin{equation*}
C_{ij}=\sum_{t} c_{i}(t)c_{j}(t)
\end{equation*}
and the sequential activation matrix is
 \begin{equation*}
S_{ij}=\sum_{t} c_{i}(t)c_{j}(t-1)
\end{equation*}
It should be noted that different normalizations of these quantities can be envisioned (see \cite{Garcia2012} for a detailed discussion). 

For all the cases, where the SER model was used, we simulated $N_{R}$ = 10000 runs of t$_{max}$ = 10 (unit timestep) with randomly generated initial conditions with $6\%$ of the nodes to be in the $E$ state and the rest to be in $S$ or $R$ state with an equiprobability.
The information for the FC matrices was accumulated by taking initially the sum over the time of each matrix and then, by taking the sum over the multiple runs. The SC/FC correlations were computed with the Pearson correlation between the flattened adjacency and the co-activation / sequential activation matrix. The final average value was computed as the mean of the correlations from the 10 different initial graphs and the errors, as the standard deviation of these correlation values. We obtain the main results using the recovery probability $p$ = 0.1 and transmission probability $f$ = 0.001. 

\subsection{Phase oscillators}
The second, also well studied, model studied here is the Kuramoto model \cite{kuramoto1975self,acebron2005kuramoto}. It describes the behaviour of a large set of coupled phase oscillators and their transition to synchronisation. We use it here in a variant, where the oscillators are coupled according to the architecture of a given network \cite{rodrigues2016kuramoto}. Each of the oscillators has an intrinsic natural frequency (or 'eigenfrequency') $\omega_{i}$ and all of them are equally coupled with their neighbours with coupling $k$. The evolution of the phase of a node in a population of $N$ oscillators is governed by the following dynamics:

\begin{equation*}
\dfrac{d\theta_{i}}{dt}= \omega_{i} + \dfrac{k}{N}\sum_{j=1}^{N}A_{ij}sin(\theta_{j}-\theta_{i}), \hspace{1cm} i=1....N
\end{equation*}

This model has been instrumental in the past for understanding, how network topology determines synchronizability \cite{arenas2008synchronization} and how synchronisation patterns emerge from architectural features of networks \cite{arenas2006synchronization}. 

Investigating the behaviour of the two classes of FC, in this model, requires oscillators that have not reached the total synchronisation, which indicates the absolute 'win' of the co-activation. Thus, Gaussian noise, scaled by amplitude $\sigma$, was added, in order to delay the synchronisation process.

\begin{equation*}
\dfrac{d\theta_{i}}{dt}= \omega_{i} + \dfrac{k}{N}\sum_{j=1}^{N}A_{ij}sin(\theta_{j}-\theta_{i})+ \sigma u, \hspace{1cm} i=1....N
\end{equation*}

The matrix of functional connectivity, in this case, is constructed from the correlation coefficient between the time series of the \textit{effective frequency}:
\begin{equation}
C_{ij}(\delta t)= corr_t(\Omega_{i}(t), \Omega_{j}(t+\delta t)), \label{phaseFunctional}
\end{equation}
where
\begin{equation*} 
\Omega_{i}(t)= \langle \Delta \theta_{i}(t) \rangle_t = \dfrac{1}{2\Delta t} \sum_{t'=t-\Delta t}^{t+\Delta t-1} \theta_{i}(t'+1)-\theta_{i}(t')
\end{equation*}
for some suitable choice of a time window $\Delta t$. 

For a continuous model, such as the coupled phase oscillators, the definition of the two classes of functional connectivity is not possible in a parameter-free manner. In Eq.(\ref{phaseFunctional}) for $\delta t=0$ we have strict co-activation and with increasing time lag $\delta t$ a transition from correlations dominated by co-activation to correlations dominated by sequential activation (before the two timeseries of effective frequencies essentially de-couple. Particularly, the decision of the appropriate selection of the time lag for the sequential activation was based on the results of SC/FC correlations as a function of the coupling strength for different values of time lag. Figure \ref{figS4} shows the multiple curves of the different time delay values for a modular and an ER graph. While the effect of the increasing time delay in a modular graph is the gradual decrease of the SC/FC correlation, in the ER graph three groups of curves emerge: the first one corresponds to the co-activity of the nodes (includes the zero and time lag equal to 1), the second group includes the curve that corresponds to the time-delay 2 and, in this case, is the appropriate selection for the sequential activation, since larger values for the time delay, that constitutes the third group of curves have zero contribution in the sequential activation.

For this case, we simulated $N_{R}$ = 100 runs over t$_{max}$ = 50, using the Euler method, with randomly generated initial conditions from the uniform distribution $(-\pi, \pi)$ on different graphs with non-identical oscillators. The integration timestep for the solution of the system was equal to 0.1. The Gaussian noise was selected to have zero mean, unit variance and it was scaled by amplitude $\sigma = 0.25$. The eigenfrequencies were uniformly selected from the interval (0,1). The size of the time window we selected $\Delta t$ for the effective frequency was equal to 20 and the FC matrices were constructed from the Pearson correlation of the effective frequencies between each pair of nodes. The diagonal elements are zero, by default. As in the SER model, the SC/FC correlations were computed with the Pearson correlation of the flattened adjacency and FC matrices. For the later one, the sum, over multiple runs, was taken and the average correlation values derived from the SC/FC correlations of 10 different initial networks; the corresponding errors derived from the standard deviation of these ten values. For the main results, we selected the coupling strength equal to 10.

\subsection{Logistic map}
The third model that was used as a dynamical probe of network architectures is the logistic map. Such dimensional maps (also termed finite-difference equations or recursion relations) are used to describe the evolution of one variable over discrete steps in time, following a template of the form $x_{t+1}=f(x_{t})$. 
The logistic map 
\begin{equation*}
x_{t+1}=R x_{t}(1-x_{t})
\end{equation*}
is the most well-known example of this class of dynamical models \cite{may1976simple}. Starting from a stable fixed point at low $R$ the system undergoes a sequence of period-doubling bifurcations with increasing $R$ leading to a large regime of deterministic chaos, occasionally interrupted by small periodic windows. Systems of coupled logistic maps have been studied extensively as a model for spatiotemporal pattern formation \cite{vadivasova2016correlation} and on networks \cite{lind2004coherence,masoller2011complex, rubinov2009symbiotic}.

The coupled system has the form
\begin{equation*}
x_{i}(t+1)=R_{i}x_{i}(t)(1-x_{i}(t))+ \dfrac{k}{N}\sum_{j=1}^{N}A_{ij}(x_{j}(t)-x_i(t)) , \hspace{0.3cm} i=1....N , 
\end{equation*}
where $k$ is the coupling strength and $A_{ij}$ is the network's adjacency matrix (structural connectivity). 
Note that we impose additional constraints on the system to force each $x_i(t)$ to be in the interval $x \in [0,1]$. 
We define FC as the correlation between the timeseries of the nodes for zero time lag (co-activation) and a time lag of 1 (sequential activation): 
\begin{equation*}
C_{ij}=corr_t(x_{i}(t), x_{j}(t)), \; S_{ij}=corr_t(x_{i}(t), x_{j}(t+1)).
\end{equation*}

We simulated $N_{R}$ = 50 runs over t$_{max}$ = 500 (unit timestep) with randomly generated initial conditions from the uniform distribution (0,1). The parameter $R$ is randomly selected by each oscillator form the interval (3.7, 3.9). For the main results the coupling strength that was used was equal to 2.
The FC matrices were constructed from the Pearson correlation between the time series of the $x$ variable (diagonal elements are zero by default). The SC/FC correlations derived from the comparison of the flattened adjacency and FC matrices, by taking the Pearson correlation, after each run. The average correlation value derived from the mean correlation values over the multiple runs and the errors from the standard deviation of the correlation values over the multiple runs.

\subsection{FitzHugh-Nagumo model}
As a more sophisticated model of excitable dynamics and regular oscillations we use the FitzHugh-Nagumo model \cite{fitzhugh1961impulses,nagumo1962active}, a 2-dimensional model of ordinary differential equations (ODEs). 

The FitzHugh-Nagumo model is composed of two coupled variables, where $x$ represents the membrane potential and $y$ is the recovery variable:
\begin{eqnarray*}
&& \tau_{x}\dfrac{\partial x_{i}(t)}{\partial t}=\gamma x_{i}(t)-\dfrac{x_{i}^{3}(t)}{3}-y_{i}(t)+ \dfrac{k}{\langle d \rangle} \sum_{j} A_{ij}[x_{j}(t)-x_{i}(t)] + \sigma v_{x} \\ 
&& \tau_{y} \dfrac{\partial y_{i}(t)}{\partial t} = x_{i}(t)-\beta y_{i}(t)+a ,
\end{eqnarray*}
where $\langle d \rangle$ is the average degree in the network, $\tau_{x}, \tau_{y}$ are the time scale parameters for each variable, again $k$ the coupling strength among the connected nodes, $v_{x}, v_{y}$ are random variables drawn from a Gaussian distribution of zero mean and unit variance and $\sigma$ the amplitude of the noise. In the $xy$ plane we can distinguish three regions and the intersection of the nullclines of the system (see Supplementary Figure \ref{supFig0}), $\dfrac{\partial x_{i}(t)}{\partial t}=0 \wedge \dfrac{\partial y_{i}(t)}{\partial t}=0$ defines the fixed point. Hence, depending on the region that the fixed point is placed, the system can be found either in the oscillatory or in the excitable regime. By shifting the linear nullcline (changing the parameter a), we can move from region 1 (excitable regime) to the oscillatory regime (region 2).
Here, we plot the correlation values during the randomisation process of a modular graph in the excitable and in the oscillatory regime.

As with the logistic map, coupled FitzHugh-Nagumo oscillators have been employed in a range of investigations focusing on spatiotemporal pattern formation \cite{grace2013predictability} and collective dynamics in networks \cite{messe2015closer}.

We simulated 10 runs, using the Euler method to solve the system. The total time of each simulated run was 180s and the integration step 0.1 ms. We downsampled the output at 1ms and we used this to calculate the FC. The FC$_{\mbox{{\scriptsize sim}}}$ matrix derived from the sum of the co-activation matrices over the time of each run. The co-activation matrices were constructed as in the SER model, after discretising the time series (spike detection) with a threshold equal to one and using a time window equal to 1 ms. For the FC$_{\mbox{{\scriptsize seq}}}$ matrices, various widths of time windows were selected in order to discretize the time series and detect the spikes. Larger time windows include both spikes that occur simultaneously and sequentially, thus, from the whole activity within the window, the co-activity (time window 1 ms) was subtracted. The calculation of SC/FC correlations derived from the flattening adjacency and functional connectivity matrices, after excluding the diagonal elements. The final correlations values came from the mean value of the 10 correlation values from the different runs and the errors from the corresponding standard deviation. The co-activity of the nodes, as well as, the sequential activity of the nodes, using different window sizes, were tested under different values for the coupling strength and the noise amplitude for both the excitable ($a=0.8$) and oscillatory regime ($a=0$) (see \ref{figS5}). The selected parameter values for the system are $\beta$ = 0.6, $\gamma=1$, $\tau_{x}$ = 0.001, $\tau_{y}$ = 0.1. The random numbers for the noise $u_{x}$ were selected from a normal distribution with zero mean and unit variance, whose amplitudes were scaled by $\sigma$ and with an additional scaling parameter $\sqrt{\dfrac{dt}{\tau_{x}}}$ (dt is the size of the integration step). The scaling term for the $u_{y}$ is equal to zero. For the main results, we selected the coupling strength (divided by the average degree in the network) equal to 0.044, the amplitude of the noise equal to $\sigma$=0.15. and the time window of 12 ms for the sequential activation.

\section*{Author contributions}
VV and MTH designed research. VV and AM wrote computational code and performed simulations. VV and MTH analysed results and wrote the framework of the paper. LJB, JC, MG, AJP, JP, RP, ST, LT and JW  wrote the geomorphology section. AF, TH and SR wrote the freshwater ecology section. MTH wrote the systems biology section. DB, AB and VL wrote the neuroscience section. MDM, BF, CKer, CKim, YS and HW wrote the social-ecological systems section. CP wrote the SNA section and contributed to the application section. All authors read and approved the final version of the paper. 

\section*{Acknowledgements}
This project has received funding from the European Union's Horizon 2020 research and innovation programme under the Marie Sk\l{}odowska-Curie grant agreement No 859937.

\vskip1pc










\vskip2pc



\clearpage

\bibliographystyle{nar}
\bibliography{sc_fc_review_2021_v2_new, additional_refs_SC_FC_temp1, neuroscience_refs, SES_refs, WP3}

\clearpage
\pagebreak

\section*{Supporting information}

\subsection*{Structural connectivity of the real life networks}
The macaque cortical area network is derived from measurements of inter-areal connection strengths (determined via labelling of neurons, \cite{markov2014weighted}) and thus can be thought of as the 'hardware', on which activity patterns emerge. The metabolic network represented in the second column of Figure \ref{fig0} is derived from the set of biochemical reactions, which are catalysed by enzymes and form the core of the metabolism of \textit{E. coli}. In fact, this figure shows the metabolite-centric projection of the bipartite metabolite-enzyme network. For each organism, the metabolic network is determined by the enzymes encoded in the organism's genome. Again, such a network represents the 'hardware' (or 'chemical space'), on which metabolic dynamics take place. 

For the social network represented in the third column of Figure \ref{fig0} the situation is less clear. The network is derived from the responses of a skills awareness questionnaire and an edge from node $A$ and $B$ represents the knowledge $A$ has of the skills of $B$ within a company \cite{cross2004hidden}. 
This information is here represented as an undirected graph, to emphasize that such a perception of competence creates a link between two individuals. An alternative would be to consider this network as a directed graph. Similar arguments could be made for the other two networks: The cortical area network is a weighted graph and the weight matrix is almost, but not completely, symmetric. Hence, not all edges are bidirectional. For Figure \ref{fig0} we used only the strongest interactions and considered them as bidirectional. In metabolism many biochemical reactions, at a given temperature and pH, have a preferential direction. As the purpose of our investigation is to illustrate, how the contributions of the two types of functional connectivity to SC/FC relationships emerge from network topology, coupling strength and dynamical parameters, we opted here to avoid such additional complications and hence for a focus on undirected graphs. Returning to the nature of the social network represented in Figure \ref{fig0}, though the perception of competences will change over time, we opted here to think of this network as structural connectivity, on which the everyday dynamics and the various types of information flow in the company take place, as the dynamics of information exchange in a company take place on a much faster time scale than the slow evolution of competences. 

\subsection*{Social Network Analysis (SNA)}
Social network analysis adopts concepts and techniques from graph theory, matrix algebra, and stochastic models to consider how structural patterns (and their interdependencies) either shape or co-evolve with particular behaviours of individual nodes or the network as a whole \cite{prell2012social}. Stochastic modelling environments such as Exponential Random Graph Models, ERGMs \cite{robins2007introduction} and Stochastic Actor-Oriented Models, SAOMs \cite{snijders2010introduction} have been developed to handle the interdependent nature of network data, and enable analysts to test theoretically-driven hypotheses while controlling for a number of competing, endogenous (or other) tendencies. These modelling suites can handle multi-relational, one-mode, bipartite, and/or multileveled network data. For network (or co-evolutionary) dynamics, the SAOMs are more commonly used. Here, an analyst considers whether social actors (or non-human entities, e.g. fish) change their behaviours in response to their position in a particular network structure(s) or set of network structures, or whether these behaviours predict particular structural patterns within or across the network(s). Here, network structure is operationalized via micro configurations, such as the presence of reciprocal ties, closed triads, or open 2-stars. An analysis models whether the presence of a given configuration (whilst controlling for other ones) predicts changes in behaviour, or vice versa (as is the case of co-evolution). In specifying configurations, a social network analyst turns to social theories for guidance, and seeks to operationalize aspects of a given theory via the choice and combination of specific configurations. These models are then typically built in a step-wise fashion, beginning with a few simple configurations as specifications (e.g. reciprocial ties), then increasing the level of complexity to include higher-order configurations that include actor attributes and/or ties from different network levels (e.g. reciprocated communication ties between actors from different organizations). Model building only ceases once the distribution of estimated, specified models ‘fit’ the observed model, according to various auxiliary statistics.

\pagebreak

\subsection*{Supporting figures}

\setcounter{figure}{0}
\renewcommand{\thefigure}{S\arabic{figure}}

\setcounter{table}{0}
\renewcommand{\thetable}{S\arabic{table}}

\begin{figure}[!h]
	\centering
	\includegraphics[width=  \linewidth]{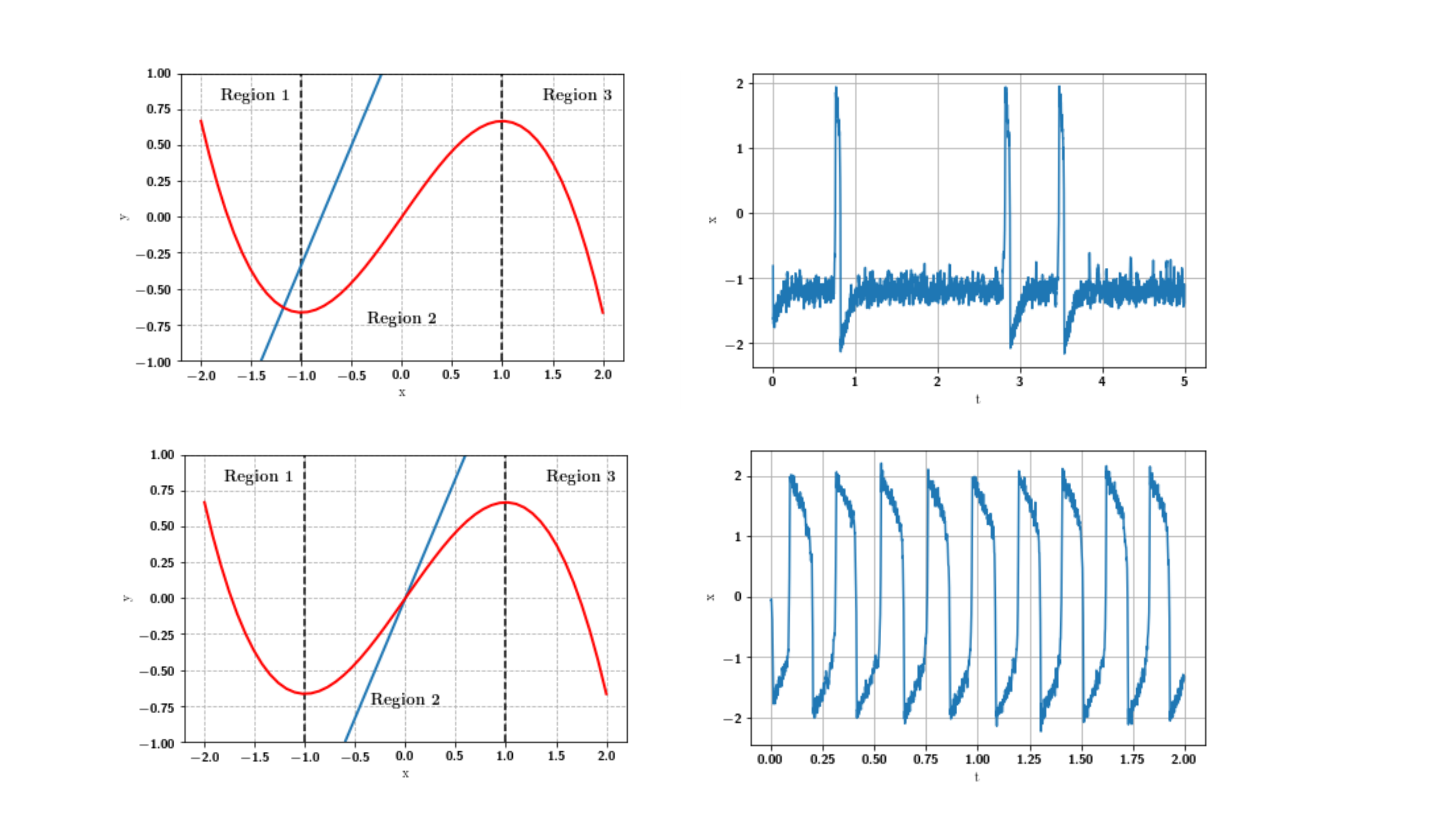}
	\caption{Phase plane and examples of time courses for the FitzHugh-Nagumo system.}
	\label{supFig0}
\end{figure}

\begin{figure}[!h]
	\centering
	\includegraphics[width=  \linewidth]{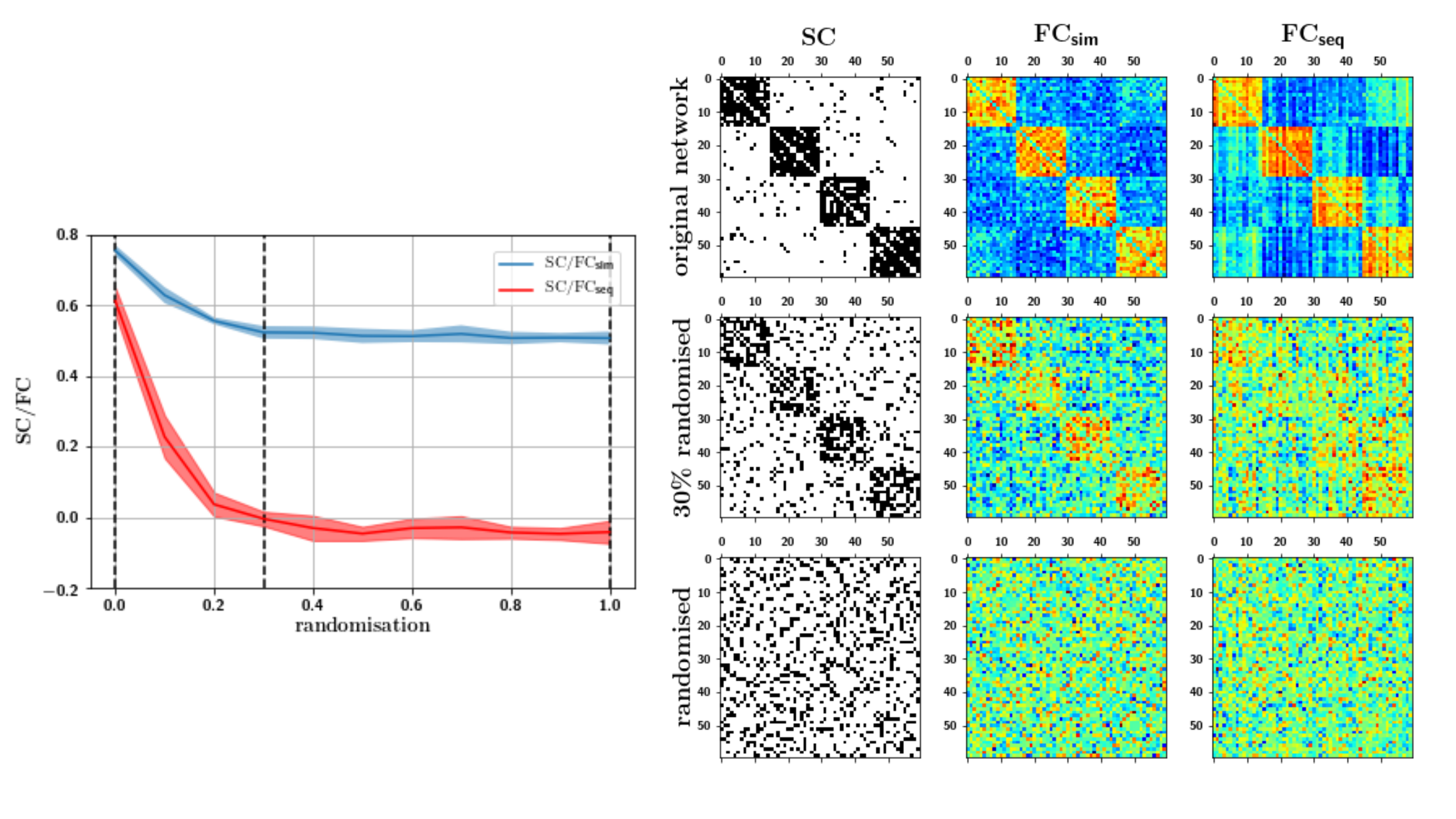}
	\caption{Same as Figure \ref{fig2}, but for coupled phase oscillators.}
	\label{figS1}
\end{figure}

\begin{figure}[!h]
	\centering
	\includegraphics[width=  \linewidth]{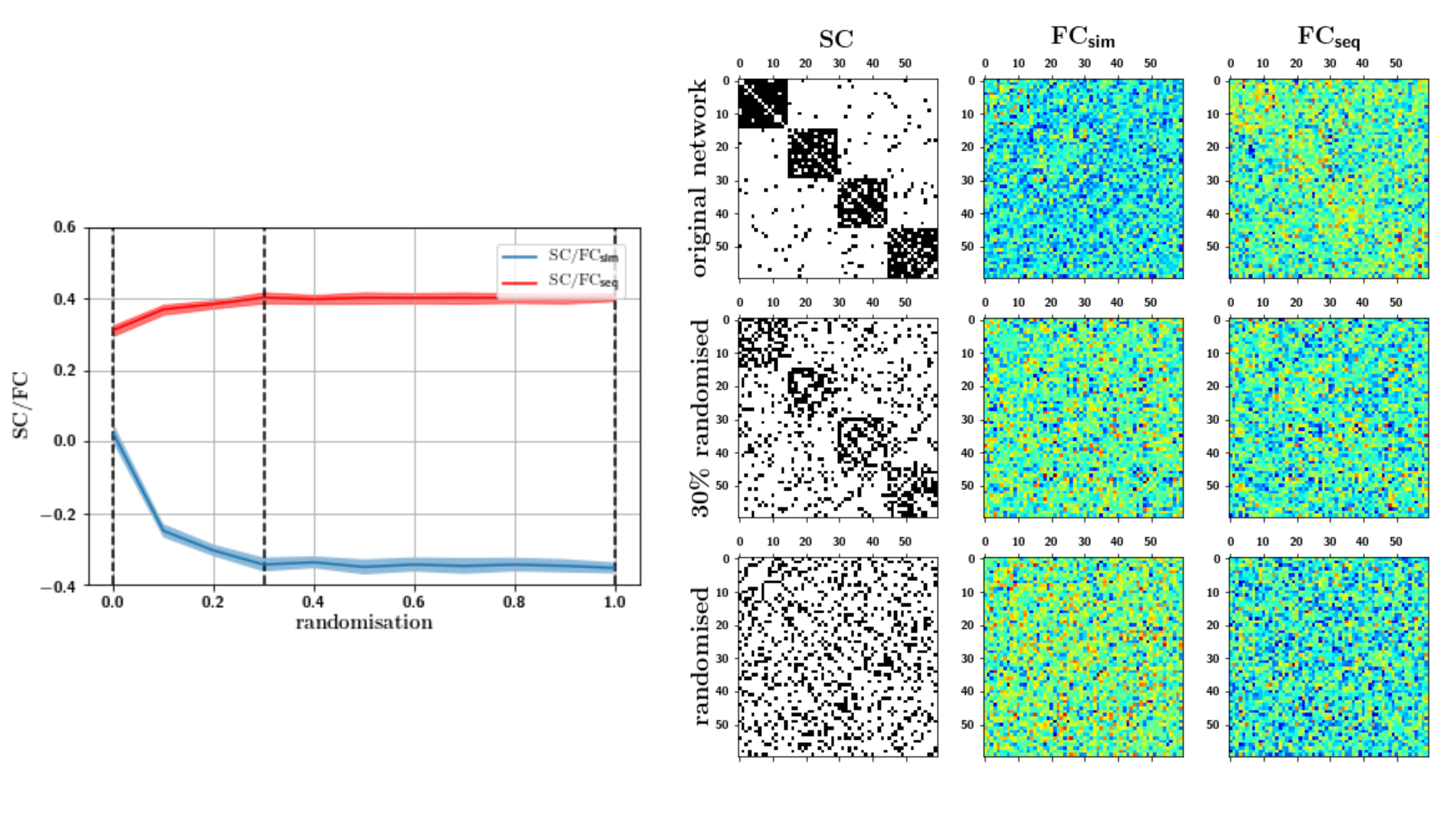}
	\caption{Same as Figure \ref{fig2}, but for the logistic map.
}
	\label{figS2}
\end{figure}

\begin{figure}[!h]
	\includegraphics[width=  \linewidth]{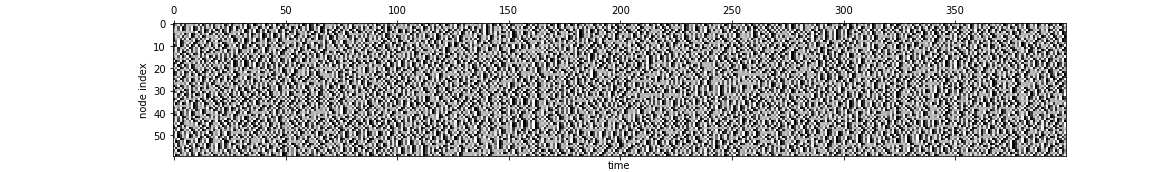} 
	\includegraphics[width=  \linewidth]{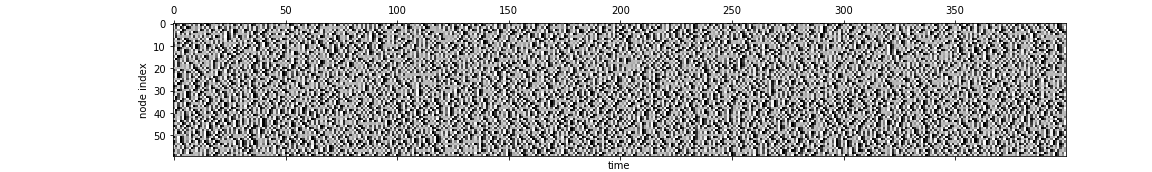} 
	\caption{Examples of space-time plots for the SC/FC results in Figure \ref{figS2} (randomisation of a modular graph under chaotic dynamics). Nodes are arranged sequentially from top to bottom, time is shown from left to right, and dynamical states are encoded in grayscale. \textbf{Top:} modular graph. \textbf{Bottom:} randomised (ER) graph.
}
	\label{figS3}
\end{figure}

\begin{figure}[!h]
	\includegraphics[width= 0.5 \linewidth]{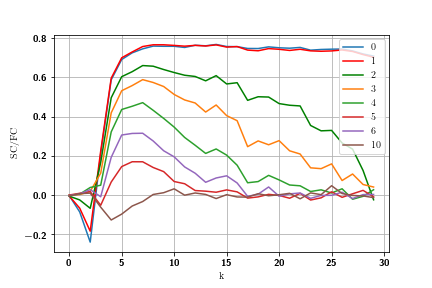} 
	\includegraphics[width=  0.5\linewidth]{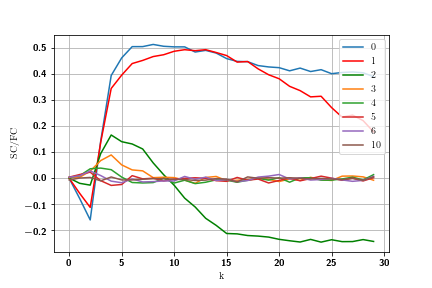} 
	\caption{Examples of SC/FC correlations derived from the coupled phase oscillators as in Figure \ref{fig4} for multiple time delays.\textbf{ Left:} modular graph (gradually reduced correlation values with the time-delay). \textbf{Right:} randomised (ER) graph (emergence of three groups of curves). 
}
	\label{figS4}
\end{figure}

\begin{figure}[!h]
	\includegraphics[width= \linewidth]{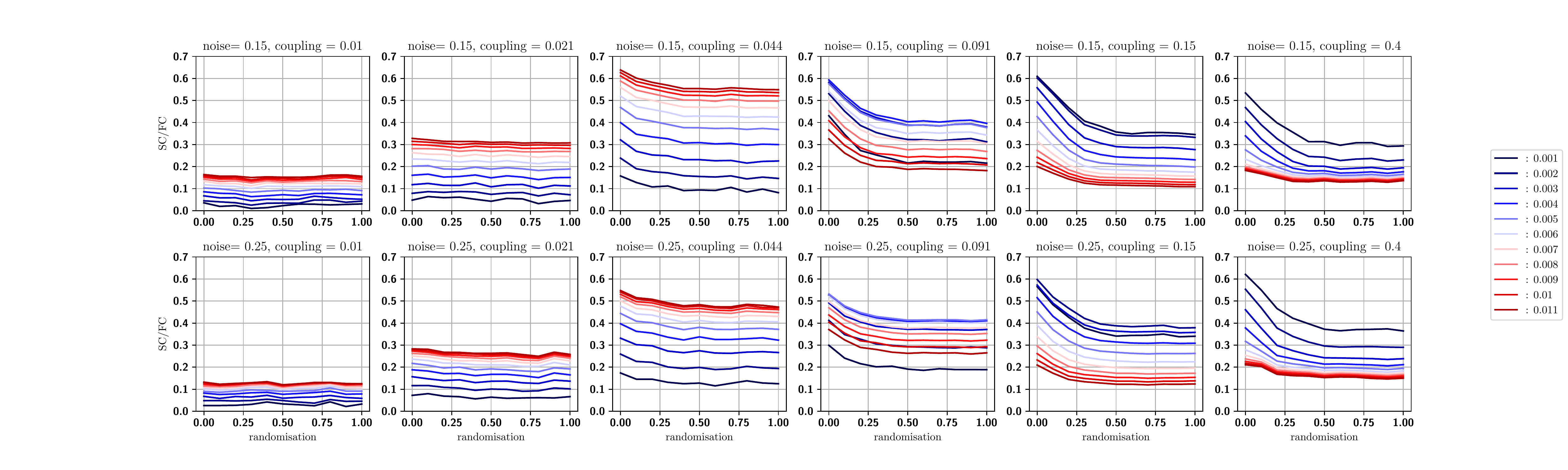} 
	\includegraphics[width=\linewidth]{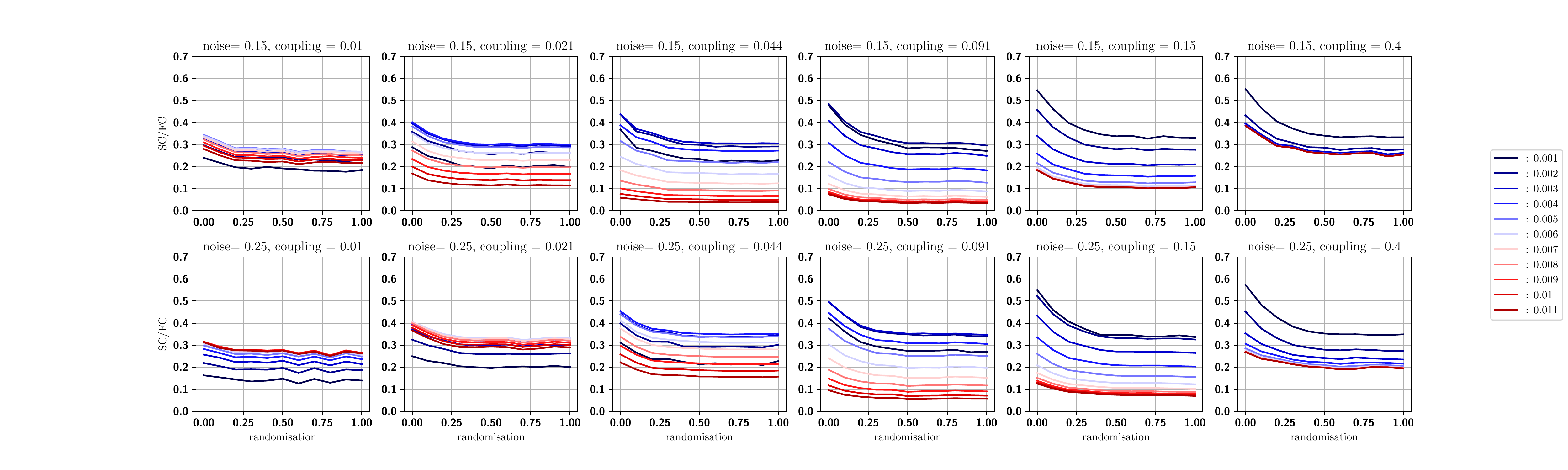} 
	\caption{Examples of SC/FC correlations as in Figure \ref{fig6} for multiple coupling strengths, noise amplitudes and time windows.\textbf{ Top:} excitable regime. \textbf{Bottom:} oscillatory regime. 
}
	\label{figS5}
\end{figure}

\end{document}